\documentclass{aa}  
\usepackage[colorlinks=true, linkcolor=blue, citecolor=blue]{hyperref}
\usepackage{graphicx}
\usepackage{txfonts}
\usepackage{longtable}
\usepackage{lipsum}
\usepackage{braket}
\usepackage{subcaption}        
\usepackage{lscape}             
\usepackage{placeins}           
\usepackage{natbib}
\usepackage{xcolor}
\usepackage{multicol}
\usepackage{float}
\newcommand{\kms}{km s$^{-1}$}
\newcommand{\civ}{\ion{C}{IV}}
\newcommand{\ciii}{\ion{C}{III}}
\newcommand{\cii}{\ion{C}{II}}
\newcommand{\mgii}{\ion{Mg}{II}}
\newcommand{\feii}{\ion{Fe}{II}}
\newcommand{\feiii}{\ion{Fe}{III}}
\newcommand{\alii}{\ion{Al}{II}}
\newcommand{\oi}{\ion{O}{I}}
\newcommand{\oii}{\ion{O}{II}}
\newcommand{\oiii}{\ion{O}{III}}
\newcommand{\siiv}{\ion{Si}{IV}}
\newcommand{\siii}{\ion{Si}{II}}
\newcommand{\oiv}{\ion{O}{IV}}
\begin{document}

   \title{The Multiphase CGM in the Epoch of Reionization: \cii\ and \civ\ absorbers around [\oiii] Emitters}
   \authorrunning{Piscitelli et al., 2026}

   \author{C. Piscitelli\inst{1,2}\corrauth{christian.piscitelli@inaf.it}        
        \and V. D'Odorico\inst{2,3}\email{valentina.dodorico@inaf.it}
        \and M. Galbiati\inst{2}
        \and M. Bischetti\inst{4,2}
        \and G. Cupani\inst{2,3}
        \and R. Dutta\inst{5}
        \and S. Cantalupo\inst{6}
        \and S. Di Stefano\inst{1,2}
        \and S.T. Guida\inst{7}
        \and K. Knudsen\inst{8}
        \and C. Mazzucchelli\inst{9}
        \and L. Paquereau\inst{8}
        \and L. Pentericci\inst{10}
        \and R. Rana\inst{8}
        \and F. Ricci\inst{11}
        \and F. Salvestrini\inst{2}
        \and A. Tortosa\inst{12,10}
        \and A. Travascio\inst{2}
        \and C. Vignali\inst{12,13}
        \and L. Zappacosta\inst{10}}

   \institute{Department of Physics, Astronomy Section, University of Trieste, Via G.B. Tiepolo, 11, I-34143 Trieste, Italy 
   \and INAF - Osservatorio Astronomico di Trieste, Via G. B. Tiepolo 11, I-34143 Trieste, Italy 
   \and IFPU - Institute for Fundamental Physics of the Universe, via Beirut 2, I-34151 Trieste, Italy 
   \and Dipartimento di Fisica “Enrico Fermi”, Università di Pisa, Largo Bruno Pontecorvo 3, Pisa I-56127, Italy 
   \and IUCAA, Postbag 4, Ganeshkhind, Pune 411007, India 
   \and Department of Physics, Università degli Studi di Milano-Bicocca, I-20126 Milano, Italy 
   \and Max-Planck-Institut für Extraterrestrische Physik (MPE), Giessenbachstrasse 1, 85748 Garching bei München, Germany 
   \and Department of Physics and Astronomy, Chalmers University of Technology, SE-412 96 Gothenburg, Sweden 
   \and Instituto de Estudios Astrof\'{\i}sicos, Facultad de Ingenier\'{\i}a y Ciencias, Universidad Diego Portales, Avenida Ejército Libertador 441, Santiago, Chile 
   \and INAF - Osservatorio Astronomico di Roma, Via Frascati 33, I-00078 Monte Porzio Catone, Italy 
   \and Department of Mathematics and Physics, Roma Tre University, Via della Vasca Navale 84, Rome, 00146, Italy 
   \and INAF - Osservatorio di Astrofisica e Scienza dello Spazio di Bologna, Via Gobetti, 93/3, I-40129 Bologna, Italy 
   \and Department of Physics and Astronomy (DIFA), University of Bologna, Via Gobetti, 93/2, I-40129 Bologna, Italy} 

  \abstract
  {We investigate the multiphase circumgalactic medium (CGM) during the Epoch of Reionization (EoR, $z>6$) by cross-correlating cool- \cii\ and warm-ionized \civ\ absorption systems with star-forming [\oiii] emitters. JWST/NIRCam wide-field slitless spectroscopy from the EIGER survey is combined with medium- and high-resolution optical/NIR spectra of six background quasars, including new VLT/X-Shooter observations of PSO J159-02. We analyze the relation between 16 \cii\ and 14 \civ\ absorbers ($\log(N) > 13.0$) and 136 galaxies, within an impact parameter of R$_{\perp} \leq$ 1000 pkpc and a line-of-sight separation of $\Delta v \leq 500$ \kms. We detect a statistically significant excess of both ions around galaxies compared to a randomized background. We observe that the \civ\ covering fraction remains enhanced up to $\sim 1$ pMpc, whereas \cii\ drops to the background level beyond $\sim 0.5$ pMpc, demonstrating that the warm-ionized phase is more spatially extended than cooler gas. Jointly, the 3D galaxy-absorber spatial clustering is significantly weaker than the galaxy-galaxy auto-correlation. This provides direct physical evidence that early carbon enrichment is not confined to the virial radius of massive star-forming systems; rather, a substantial fraction of these metals permeates the diffuse intergalactic medium (IGM) or is injected by a widespread population of faint, undetected dwarf galaxies. Finally, we note a rapid radial decline of the \civ\ covering fraction compared with lower redshift samples at $z<2$ and $z\sim3-4$ pointing out an evolving CGM ionization structure where early metals reside predominantly in lower ionization states. In conclusion, we determine a conservative lower limit for the observed carbon mass of $\rm M_{\cii + \civ} \geq 2.8 \times 10^6 \, M_{\odot}$ within 300 pkpc.}


   \maketitle

\nolinenumbers
\section{Introduction}
The circumgalactic medium (CGM) plays a fundamental role in the evolution of galaxies, serving as an interface between galaxies and the intergalactic medium (IGM). The exchange of baryons, metals, and energy is regulated through accretion of metal poor gas, outflows from feedback processes occurring inside galaxies and recycling of metal-enriched material \citep[for a review, see][]{annurev_Tumlinson2017, 2020Per_annurev}. The interplay between accretion and feedback processes regulates the availability of the fuel needed by galaxies to sustain their star formation \citep[e.g.,][]{2009_Dekel} and at the same time, it enriches and shapes the properties of the CGM gas, producing a multiphase medium that spans a wide range of temperatures, densities, kinematic properties, and chemical abundances \citep[e.g.,][]{2008_Oppenheimer, annurev_Faucher2023}. 

Due to the low density nature of CGM gas, and thus to its faintness, it is commonly studied in absorption, through the line features it imprints in the spectra of background bright sources, such as quasars. 
However, it is the combination of deep galaxy surveys and high-resolution quasar spectroscopy that has allowed the statistical characterization of the CGM at $z \lesssim 4$ \citep[e.g.,][]{2005_Adelberger, 2010_Steidel, 2012_Rudie, 2014_Turner, 2015_Fumagalli, Burchett_2016, 2017_Chen, 2019_Rudie, Dutta_2020_MAGGII, Dutta_2021, 2021_Schroetter, 2023_Galbiati, 2023_Banerjee, 2024_Galbiati, 2025_Banerjee, 2026arXiv260107916B}. These studies commonly use metal doublet transitions, such as \mgii\ $\lambda\lambda\,2798, 2803$ \AA\ and \civ\ $\lambda\lambda\,1548, 1550$ \AA, to trace the CGM gas, probing different gas phases ranging from the cool (T $\sim 10^4$ K) gas to a more ionized and, sometimes, warmer phase (T $\sim 10^5$ K). 
Observations, together with results from simulations \citep[e.g.,][]{2013_Ford, 2021_Ho, Weng_2024}, reveal that the presence and column density of neutral hydrogen and metal absorbing gas show strong statistical correlation with distance from galaxies and from the environment, with galaxy groups richer in enriched gas with respect to isolated galaxies \citep[e.g.,][]{Dutta_2021,2023_Galbiati}. Furthermore, low-ionization ions (e.g., \mgii, \cii, and \oi) are mostly concentrated closer to the galaxies than higher ions (e.g., \civ, \siiv, and \oiv), which can be detected even up to several times the galaxies’ virial radius \citep[e.g.,][]{2021_Schroetter, 2023_Qu}.

Analogous investigations in the very high-redshift Universe ($z \gtrsim 5$) are still rare and based on single fields \citep[e.g.,][]{Diaz_2021}, particularly during the Epoch of Reionization (EoR, $z>6$), when the most massive structures were in the early stages of assembly, and the interplay between galaxies and their surrounding gas and nearby galaxies was not yet fully established. In this regime, environmental interactions may be less developed than in the Local Universe. More recently, high-redshift studies based on ground- and space-based near-infrared (NIR) quasar spectroscopy have shown that the early Universe was already chemically enriched, revealing an increase in the number density of low-ionization absorbers and a decline in high-ionization species with increasing redshift \citep{Becker_2019, 2022_DOdorico, davies2023a, davies2023b, 2023_Christensen, 2024_Sebastian}. At the same time, the \textit{James Webb Space Telescope} (JWST) has opened new frontiers providing deep surveys of galaxies such as the \textit{Emission-Line Galaxies and Intergalactic Gas in the Reionization Epoch} \citep[EIGER,][]{2023_Kashino, 2023_Matthee,2026_Kashino} and \textit{A SPectroscopic survey of biased halos in the Reionization Era} \citep[ASPIRE,][]{2023_Wang_ASPIRE}, both centered on samples of $z\sim6$ quasars for which spectra were available. \citet{Bordoloi_2024} and \citet{Higginson_2026} studied the connection between high-redshift [\oiii] emitters in the EIGER fields and \mgii\ and \oi\ absorbers, respectively. They found that most of the high-$z$ \mgii\ absorption is not consistent with being bound to the dark matter halos of the host galaxies, while \oi\ at $z\sim 6$ could primarily trace extended overdensity environments rather than the CGM of individual halos. \citet{Zou_2024}, using ASPIRE data, characterized the CGM at $z\sim6-6.5$ by linking metal absorbers to surrounding [\oiii] emitters, suggesting that their chemical abundance modeling in overdense regions favors a top-heavy initial mass function, hinting at a direct contribution from Population III stars to cosmic gas enrichment at the end of Reionization.

In this work, we exploit the catalog of [\oiii] emitters in the EIGER fields \citep{2026_Kashino} and the absorption lines identified in the spectra of the six central quasars to investigate the CGM at $z \sim 6$ through the cross-correlation between galaxies and \cii\ ($\lambda\, 1334\ \AA$) and \civ\ absorbers. The use of these two ions allows us to investigate the distribution of gas in two different ionization stages, likely tracing a colder and hotter medium. 

The paper is organized as follows. In Section \ref{sec:sample}, we describe the galaxy sample from EIGER, and the absorption line sample from proprietary and literature data.  
In Section \ref{sec:analysis}, we study the spatial distribution of \cii\ and \civ\ absorption around EIGER galaxies. In Section \ref{sec:discussion}, we discuss and compare our results with the low-redshift literature samples, and we present our conclusions in Section \ref{sec:conclusion}.

Throughout this work, we adopt a flat $\Lambda \rm CDM$ cosmology with $H_0 =70 $ km s$^{-1}$ Mpc$^{-1}$, $\Omega_M = 0.27$ and $\Omega_{\Lambda} = 0.73$. Uncertainties are computed as 1$\sigma$ (68.3 $\%$) confidence level. Distances are given in proper units using the prefix "p", unless we use the prefix "c" before the unit, in which case it refers to the comoving distance.

\section{Observational samples}
\label{sec:sample}
In this section we describe the properties of galaxies and absorption systems used for the analysis. 

\subsection{[O~III] emitters}
We used the catalog of [\oiii] emitting galaxies from the EIGER survey provided by \cite{2026_Kashino}. 
EIGER (\textit{Emission-Line Galaxies and Intergalactic Gas in the Reionization Epoch}) is a JWST program (Program ID: 1243; PI: S. J. Lilly) based on NIRCam Wide Field Slitless Spectroscopy (WFSS) of six extragalactic deep fields, each one centered on a luminous quasar at $z_{\rm qso} \ge 6.0$. A complete description of the observing program can be found in \citet{2023_Kashino, 2023_Matthee, Bordoloi_2024}.
The main properties of the six quasars of the EIGER Survey (hereafter J0148, J1030, J0100, J159, J1148, J1120) are listed in Table \ref{Quasars_table}. They were selected to have medium- and high-resolution ($R\sim 5000-45,000$) spectra obtained with Magellan/FIRE \citep{2008_FIRE}, VLT/X-Shooter \citep{2011_Vernet}, Keck/MOSFIRE \citep{2010_MOSFIRE} or Keck/HIRES \citep{vogt_hires}, amounting to several hundred hours of observations, providing a unique window into the foreground absorbing gas along their line-of-sight. In addition to these spectroscopic data, the EIGER quasar fields benefit from JWST/NIRCam \citep{2023_Rieke} and HST imaging. In particular, for the JWST imaging the NIRCam F115W, F200W, and F356W filters were used. 

\begin{table*}[ht!] 
\caption{EIGER sample of quasars.} 
\label{Quasars_table} 
\centering 
\begin{tabular}{c c c c c c c}
\hline\hline 
Quasar Name & Ra & Dec & $z_{em}$ & Ref & $\Delta z_{\cii}$ & $\Delta z_{\civ}$$^a$\\ 
 & hrs & deg & \\
(1) & (2) & (3) & (4) & (5) & (6) & (7)\\
\hline 
    ULAS J0148+0600 & 01:48:37.64 & +06:00:20.06 & 5.9895 & \cite{2023_Dodorico} & 5.367-5.874 & 5.33-5.874 \\
    SDSS J1030+0524 & 10:30:27.10 & +05:24:55.0 & 6.304 & \cite{Farina_2019} & 5.653-6.183 & 5.33-6.183 \\
    SDSS J0100+2802 & 01:00:13.02 & +28:02:25.8 & 6.3269 & \cite{Venemans_2020} & 5.674-6.206 & 5.33-6.206 \\
    PSO J159-02 & 10:36:54.19 & -02:32:37.94 & 6.3809 & \cite{2018_DeCarli} & 5.723-6.259 & 5.33-6.259 \\
    SDSS J1148+5251 & 11:48:16.64 & +52:51:50.3 & 6.4189 & \cite{2005_Maiolino} & 5.758-6.296 & 5.33-6.296 \\
    ULAS J1120+0641 & 11:20:01.48 & +06:41:24.3 & 7.0848 & \cite{Venemans_2020} & 6.365-6.951 & 5.348-6.951\\
\hline 
\end{tabular}
\tablefoot{(1) : Quasar name; (2-3) R.A. and Dec. (J2000); (4) Quasar redshift based on [\cii] emission line; (5) Redshift reference; (6) Redshift window for the detection of \cii\ systems to avoid contamination from the Ly$\alpha$ forest and quasar's proximity zone; (7) Same as (6) but for \civ.\\
$^a$: The lower bound of the redshift window is set by the lowest-redshift to which JWST observations are sensitive, with the exception of J1120 for which it is determined by the Ly$\alpha$ forest limit.}
\end{table*}

Galaxies in the EIGER fields are identified as [\oiii] $\lambda\lambda\,4960, 5008$ \AA\ emitters across the redshift range $z \sim 5.33-6.97$. The [\oiii] $\lambda\,5008 \AA$ fluxes ($F_{5008}$) have been estimated to be within the range $-18< \log(F_{5008}/\rm [erg \ s^{-1} \ cm^{-2}]) < -16.5$ and  corresponding luminosities vary from $41.6 < \log(L_{5008}[\rm erg \ s^{-1}]) < 43.1$ \citep{2026_Kashino}. 
This sample includes star-forming systems spanning a broad range of stellar masses, $6.8 \leq \log(M_{\star}/M_{\odot}) \leq 9.3$, and star formation rates averaged over the most recent 50 Myr, $-1 \leq$ $\log$$(\mathrm{SFR}$/$M_{\odot}$ $ \rm yr^{-1}) \leq 1.5$ \citep{Higginson_2026}.
The full sample consists of 948 [\oiii] emitters \citep{2026_Kashino}.

\subsection{QSO absorption systems}
In order to probe the distribution of gas in the CGM of [\oiii]-emitters, we collected a sample of \cii\ and \civ\ absorption systems along the lines of sight to the six central quasars. 

The sample used in this work is composed of both literature and proprietary data. J0148, J1030 and J0100 belong to the enlarged XQR-30 sample \citep{2023_Dodorico}, for which the catalog of metal absorption lines with measured redshift, column density and Doppler parameter is provided by \cite{davies2023a}. Concerning J1148 and J1120, the absorption systems are provided by \cite{2006_Becker} and \cite{2017_Bosman}, respectively.  However, observations of J1148 were conducted with Keck/HIRES \citep{vogt_hires}, which has a high spectral resolution ($R\sim 45000$ for a slit width of 0.86$''$) and a maximum wavelength of 1 $\mu$m. This implies that \civ\ is detectable in the J1148 spectrum only in the redshift range $4.82 \lesssim z \lesssim 5.46$, minimally overlapping with the redshift range probed by the EIGER [\oiii] emitters. 

Furthermore, in this work we present a new VLT/X-Shooter spectrum for J159, improving the signal-to-noise ratio (S/N) by a factor of $\sim2$ compared to previous observations. The final spectrum has S/N~$\sim 22$ per 10 \kms\ bin measured at $\lambda \simeq$ 1285 \AA\ rest frame. The details of the data reduction and analysis of the J159 spectrum, including the identification of absorption systems, are reported in Appendix \ref{Appendix_B} and \ref{Appendix_C}. 

Following \citet{davies2023a}, we created absorption systems by merging velocity components closer than 200 \kms. 
The total column density of the system is obtained by summing the column densities of the individual components; the corresponding uncertainty on the total logarithmic column density is calculated via standard error propagation. The redshift of the absorption system is determined as the column-density-weighted mean of the redshifts of the individual components as 
\begin{equation}
    \label{weighted_mean_col_den}
    z_{\rm abs} = \dfrac{\sum_i z_i N_i}{\sum_i N_i}
\end{equation}
and we use the standard error of the weighted mean as the corresponding uncertainty. 
Finally, to avoid the proximity region of the background quasar, we exclude both [\oiii]-emitters and absorption systems located within a velocity separation of 5000 \kms\ from the quasar systemic redshift. Accordingly, the maximum redshift considered for each quasar field is calculated as:
\begin{equation}
    z_{\rm max} = (1 + z_{\rm qso})\sqrt{\dfrac{1-\beta}{1+\beta}} -1
\end{equation}
where $\beta = v/c$. Within the same framework, to prevent contamination from the Ly$\alpha$ forest in the quasar spectra, we establish a minimum redshift threshold for both [\oiii]-emitters and absorption systems, defined as:
\begin{equation}
    z_{\rm min} = (1 + z_{\rm qso}) \biggl(\dfrac{\lambda_{\rm Ly\alpha}}{\lambda_{\rm ion,rest}} \biggr) - 1
\end{equation}
where $\lambda_{\rm Ly\alpha}$ is the rest-frame wavelength of the Ly$\alpha$ transition, and $\lambda_{\rm ion,rest}$ refers to the rest-frame wavelengths of the \cii\ and \civ\ transitions. The resulting redshift ranges computed for each quasar field are reported in Table \ref{Quasars_table}.\\
To ensure the reliability of our statistical analysis and robustly account for varying spectral noise across different sightlines, the sample is limited to systems satisfying the condition $\log (N) \ + \sigma_{\log(N)} \geq 13.0 $, corresponding to a completeness threshold $\gtrsim 50$\% in the considered absorption line samples \citep{2017_Bosman,davies2023a}. 
Following these selection criteria, our final sample consists of 16 \cii\ and 14 \civ\ absorption systems. 
We list all the \cii\ and \civ\ absorbers in Table \ref{table_Carbon_absobers}.

\section{Analysis of the relation between galaxies and absorbers}
\label{sec:analysis}

We investigate the physical properties and spatial distribution of the CGM traced by \cii\ and \civ\ absorbers at $z\sim6$, establishing the line-of-sight and angular connection between the absorbers and the EIGER galaxy population.

As a first step, we compute for every galaxy the impact parameter $R_{\perp}$ measured in physical kpc (pkpc), which is the projected distance to the quasar line of sight, as 
\begin{equation}
    R_{\perp} = \theta \; \cdot D_A(z_{\rm gal})
\end{equation}
where, $\theta$ is the angular separation on the sky between the galaxy and the quasar sight-line, and $D_A$ is the angular diameter distance calculated at the redshift of the galaxy. As a second step, we compute the line-of-sight distance for every absorber-galaxy pair as a velocity offset, measured in \kms, as
\begin{equation}
    \label{eq_velocity}
    \Delta v = c \; \cdot \dfrac{(1 + z_{\rm abs})^2 - (1 + z_{\rm gal})^2}{(1 + z_{\rm abs})^2 + (1 + z_{\rm gal})^2}
\end{equation}
where $c$ is the speed of light and $z_{\rm abs}$ and $z_{\rm gal}$ are the redshifts of the absorption system and galaxy, respectively.

\begin{figure*}[ht!]
    \centering
    \begin{minipage}{0.49\hsize}
        \centering
        \includegraphics[width=\linewidth]{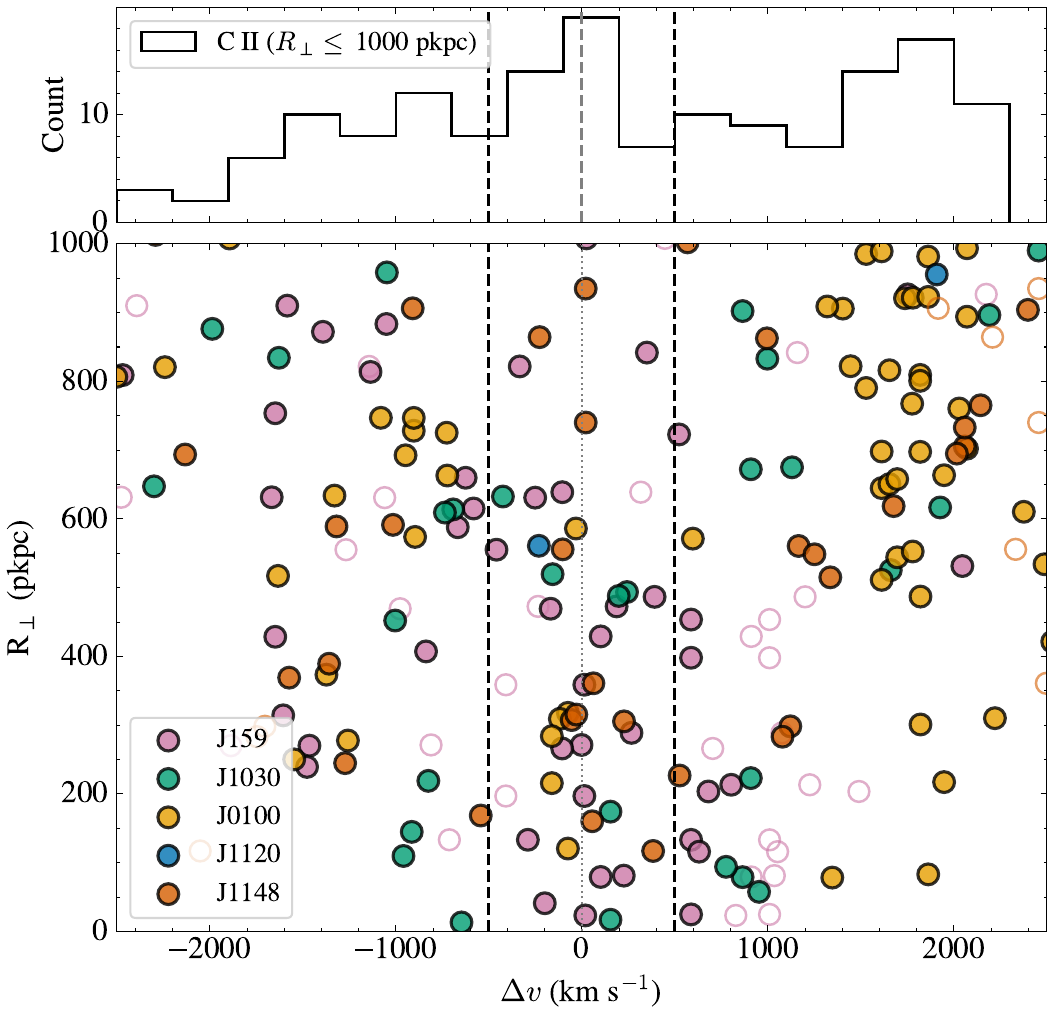}
    \end{minipage}
    \hfill
    \begin{minipage}{0.49\hsize}
        \centering
        \includegraphics[width=\linewidth]{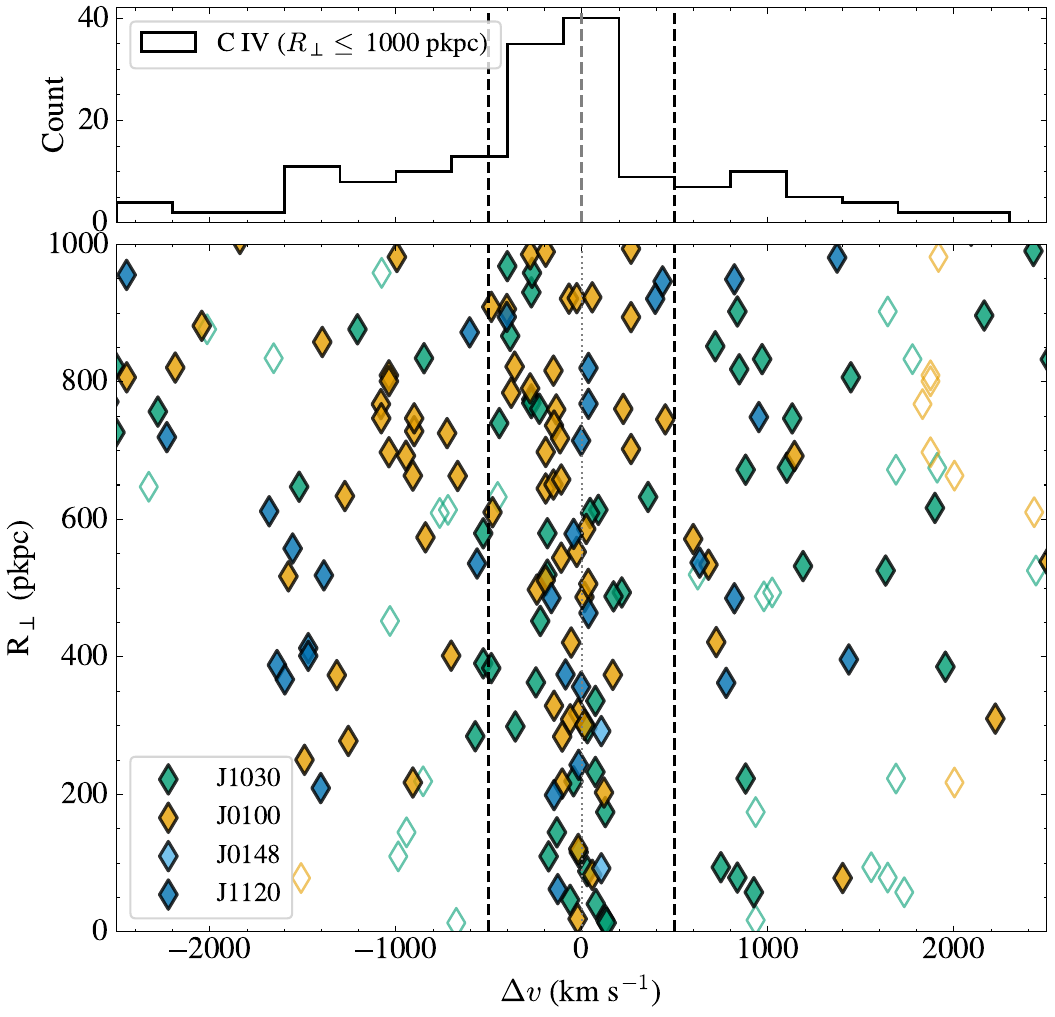}
    \end{minipage}
    \caption{Projected impact parameter versus line-of-sight velocity difference between absorbers and galaxies. The left and right panels refer to \cii\ and \civ\, respectively. The vertical dashed lines mark the $|\Delta v| \leq 500$ \kms velocity window. Filled circles and diamonds   represent primary galaxy associations, while open circles and diamonds denote secondary associations. Distinct colors indicate the different   quasar fields. The upper panels report the distribution of the number of galaxies as a function of the offset velocity. }       
    \label{kinematics_carbon}
\end{figure*}

In Fig.~\ref{kinematics_carbon}, we plot those quantities with an "absorption-centric" perspective, i.e. centering on each absorber and considering all galaxies with a velocity offset $|\Delta v| \leq 2500$ \kms, and an impact parameter $R_{\perp} \le 1000$ pkpc. 

When a single galaxy is associated with multiple absorbers at different redshifts, we define the association with the smallest velocity offset as primary, and all the others as secondary. In Fig.~\ref{kinematics_carbon}, the line-of-sight velocity distribution is obtained by including only primary associations to avoid double-counting galaxies. The distribution reveals a prominent excess of galaxies centered at $\Delta \rm v \simeq 0$ \kms. We find that the 26 \% for \cii\ and 55 \% for \civ\ of these associated systems are enclosed within a velocity difference of $|\Delta v | \le 500$ \kms (indicated by the vertical dashed lines).

Based on this result and on previous works \citep{Dutta_2021, 2023_Galbiati, 2024_Galbiati}, we adopt a velocity threshold of $\pm 500$ \kms\ to define the association between a galaxy and an absorber. Within this window and an impact parameter of R$_{\perp} \leq 1$ pMpc, we identified 44 and 92 [\oiii] emitters around \cii\ and \civ\ absorbers, respectively (see Table \ref{table_Carbon_absobers}).

\begin{figure*}[ht!]
    \centering
    \begin{minipage}{0.46\hsize}
        \centering
        \includegraphics[width=\linewidth]{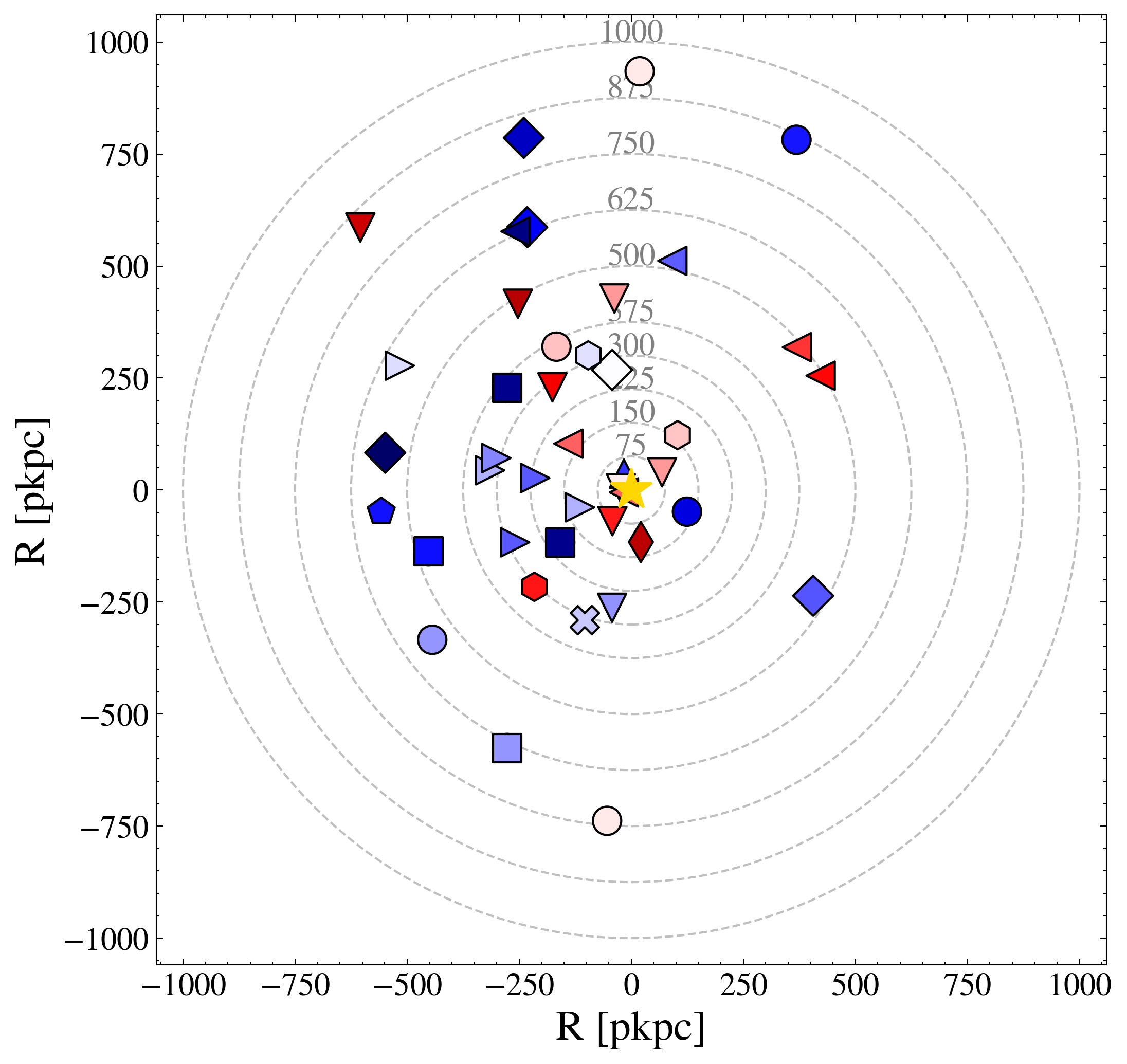}
    \end{minipage}
    \hfill
    \begin{minipage}{0.53\hsize}
        \centering
        \includegraphics[width=\linewidth]{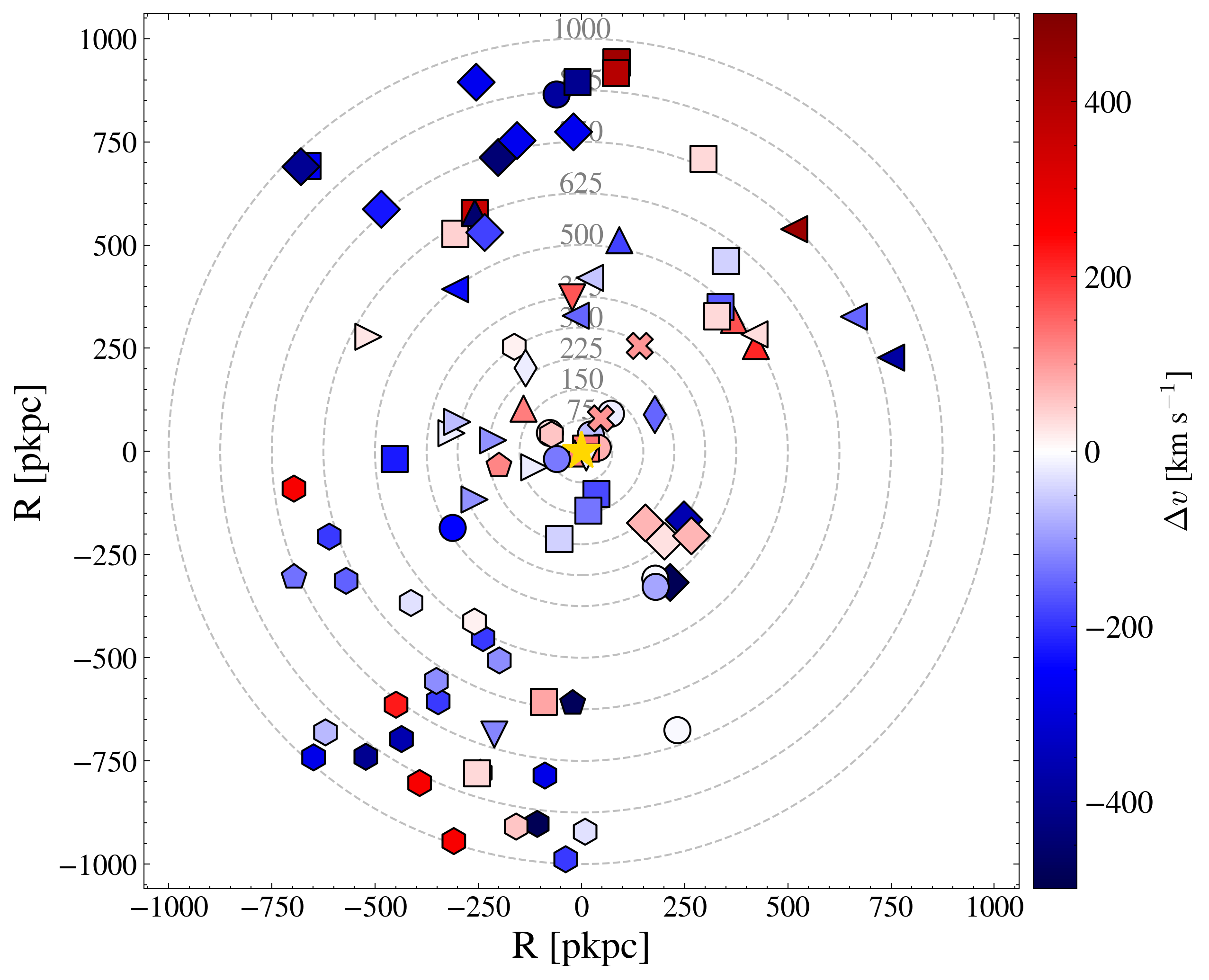}
    \end{minipage}
    \caption{2D distribution of galaxies associated with \cii\ (left panel) and \civ\ (right panel) absorbers. The central gold star indicate the position of the line of sight. The same symbol indicate different galaxies within $|\Delta v| \le 500$ \kms\ of the same absorber.}
    \label{Carbon distribution}
\end{figure*}

Finally, in Fig.~\ref{Carbon distribution} we show the 2D distribution of galaxies within $|\Delta v | \le 500$ \kms of \cii\ and \civ\ absorbers and impact parameters from the line of sight $\le 1000$ pkpc.

\subsection{Covering fraction}
\label{covering_fraction}
In order to investigate the properties of the CGM around the [\oiii]-emitting galaxies, we measure the covering fraction of \cii\ and \civ\ as a function of impact parameter. 
This quantity is defined as the ratio between the number of [\oiii] emitters having a \cii\ or \civ\ absorption system within a velocity offset $|\Delta v_{\rm max}|$ and the total number of [\oiii] emitters in the considered impact parameter bin, i.e.:   

\begin{equation}
    f_c( R_{\perp}) = \dfrac{\sum_{i \in S(R_{\perp})} A_i(\Delta v_{\rm max})}{\sum_{i \in S(R_{\perp})} 1} 
\end{equation}

\noindent
where $S(R_{\perp})$ is the sample of galaxies that fall in the chosen impact parameter bin, while $A_i (\Delta v_{\rm max})$ is a function that is equal to 1 if there exists an absorber with $|\Delta v| \leq \Delta v_{\rm max}$, 0 otherwise. In this work, we assume $|\Delta v_{\rm max}|=500$ \kms and impact parameter bins of $\Delta R_{\perp} = 75$ pkpc up to 375 pkpc, and $\Delta R_{\perp} = 125$ pkpc between 375 pkpc and 1 pMpc. The uncertainty in the covering fraction is computed by adopting the Wilson-score technique \citep{Wilson_score, Dean_2015}, which, unlike the standard normal approximation, is a method that properly accounts for the asymmetry of binomial errors in the small-sample regime, keeping the boundaries strictly confined to [0,1]. 
Results of this calculation for both \cii\ and \civ\ are reported in Table \ref{table_covering_fraction_500}. 

\begin{table*}[ht!]
\small
\caption{Differential covering fraction of \cii\ and \civ\ absorbers around [\oiii] emitters, evaluated within a velocity window of $\rm |\Delta v | \leq 500$ \kms. For completeness, we also report the \cii\ + \civ\ differential covering fraction shown in Fig.~\ref{fig:cii_plus_civ_cov_frac}}.           
\label{table_covering_fraction_500}    
\centering                        
\begin{tabular}{c c c c c c c c c c c}       
\hline\hline        
 & & \cii & & & \civ & & & \cii\ + \civ & \\ 
$\rm \Delta R_{\perp}$ & Tot & Det & $f_c$ & Tot & Det & $f_c$ & Tot & Det & $f_c$\\        
pkpc & & & \\
\hline
   0-75 & 6 & 3 & 0.500$^{+0.189}_{-0.189}$ & 10 & 6 & 0.600$^{+0.139}_{-0.157}$ & 6 & 4 & 0.667$^{+0.156}_{-0.204}$ \\[1.25ex]
   75-150 & 17 & 5 & 0.294$^{+0.097}_{-0.119}$ & 25 & 7 & 0.280$^{+0.097}_{-0.080}$ & 17 & 9 & 0.471$^{+0.119}_{-0.116}$ \\[1.25ex]
   150-225 & 17 & 4 & 0.235$^{+0.086}_{-0.116}$ & 25 & 5 & 0.200$^{+0.091}_{-0.068}$ & 17 & 6 & 0.294$^{+0.119}_{-0.097}$ \\[1.25ex]
   225-300 & 20 & 4 & 0.200$^{+0.074}_{-0.103}$ & 31 & 6 & 0.194$^{+0.080}_{-0.061}$ & 20 & 8 & 0.350$^{+0.111}_{-0.097}$ \\[1.25ex]
   300-375 & 20 & 7 & 0.350$^{+0.097}_{-0.111}$ & 41 & 9 & 0.220$^{+0.071}_{-0.058}$ & 20 & 9 & 0.350$^{+0.111}_{-0.097}$ \\[1.25ex]
   375-500 & 30 & 6 & 0.200$^{+0.063}_{-0.082}$ & 69 & 9 & 0.130$^{+0.046}_{-0.035}$ & 30 & 11 & 0.367$^{+0.091}_{-0.082}$ \\[1.25ex]
   500-625 & 46 & 5 & 0.109$^{+0.038}_{-0.054}$ & 71 & 11 & 0.155$^{+0.048}_{-0.038}$ & 46 & 14 & 0.283$^{+0.070}_{-0.061}$ \\[1.25ex]
   625-750 & 45 & 4 & 0.089$^{+0.034}_{-0.052}$ & 83 & 11 & 0.133$^{+0.042}_{-0.033}$ & 45 & 10 & 0.222$^{+0.068}_{-0.056}$ \\[1.25ex]
   750-875 & 38 & 3 & 0.079$^{+0.034}_{-0.055}$ & 63 & 12 & 0.190$^{+0.054}_{-0.044}$ & 38 & 14 & 0.368$^{+0.081}_{-0.074}$ \\[1.25ex]
   875-1000 & 36 & 1 & 0.028$^{+0.017}_{-0.028}$ & 61 & 15 & 0.246$^{+0.059}_{-0.051}$ & 36 & 14 & 0.389$^{+0.083}_{-0.077}$ \\[1.25ex]
\hline  
\end{tabular}
\tablefoot{The first column indicates the upper boundary of each 75 pkpc concentric radial bin up to 375 pkpc, after which we assume 125 pkpc concentric radial bin. The subsequent columns report the total number of galaxies surveyed in that annulus, the number of successful pairs detections, and the resulting covering fraction.}
\end{table*}

\begin{figure*}[ht!]
    \centering
    \begin{minipage}{0.49\hsize}
        \centering
        \includegraphics[width=\linewidth]{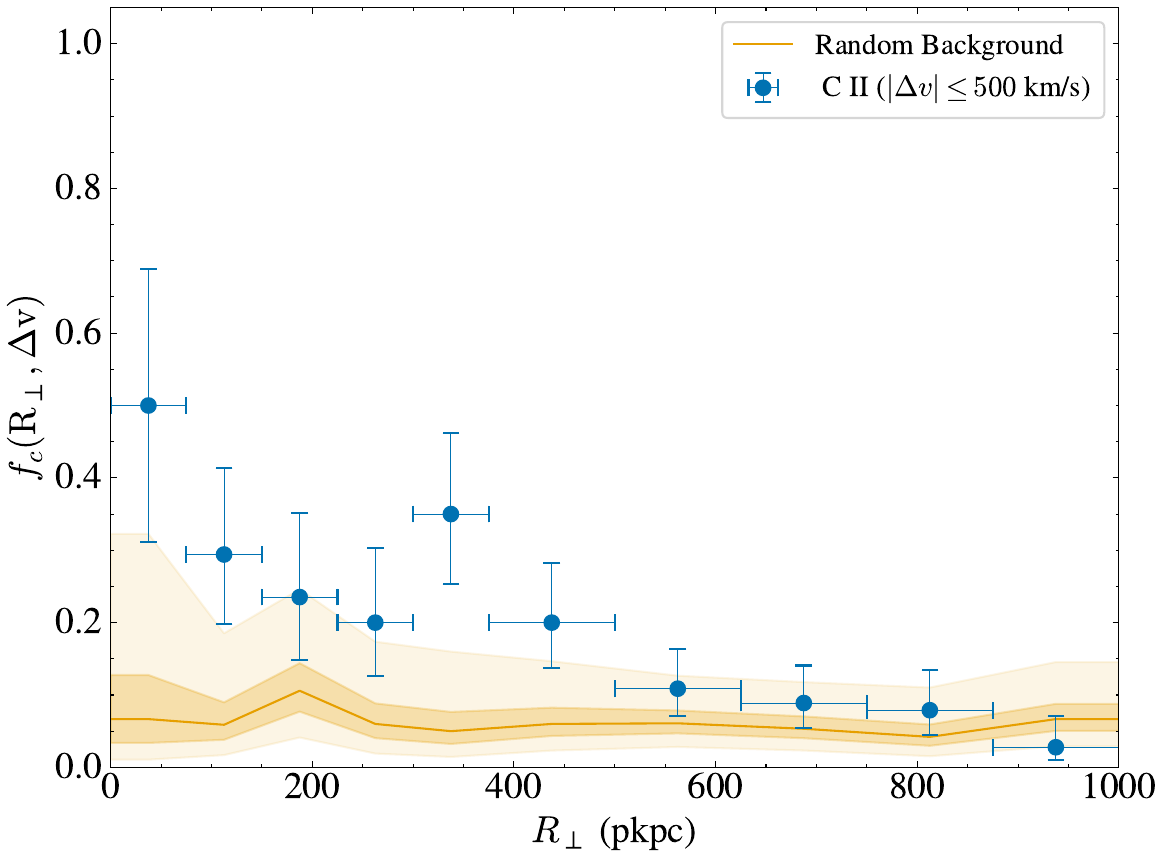}
    \end{minipage}
    \hfill
    \begin{minipage}{0.49\hsize}
        \centering
        \includegraphics[width=\linewidth]{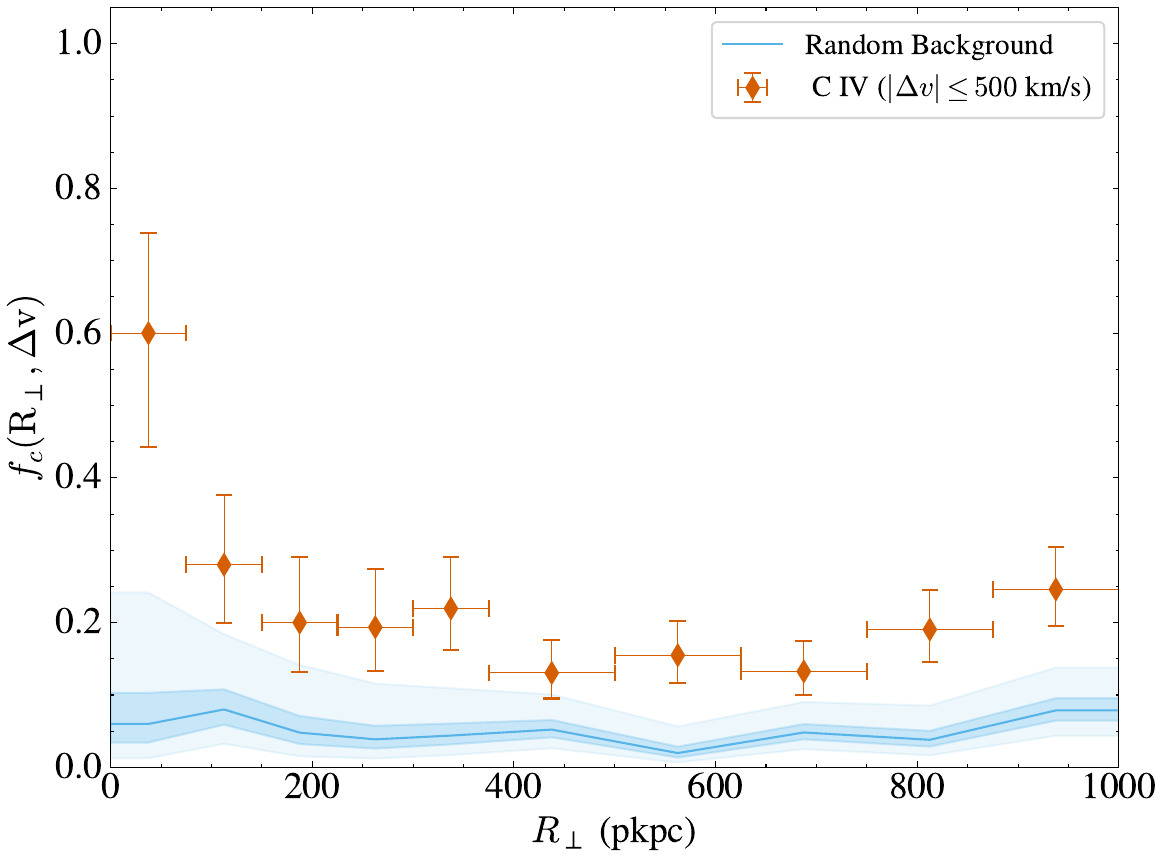}
    \end{minipage}
    \caption{Differential covering fraction of \cii\ (left panel) and \civ\ (right panel) absorbers around the surveyed galaxies. The points represent the measured signal in concentric annuli. The vertical error bars denote the Wilson-score confidence interval of 1$\sigma$, while the horizontal error bars indicate the width of each radial bin. The solid line denotes the expected random background, and the lighter and darker shaded regions indicate the 1$\sigma$ and 3$\sigma$ confidence intervals, respectively.}
    \label{dcf_500_75}
\end{figure*}

To evaluate the significance of the measured covering fractions, we compare them with a background expectation.
To this end, we employ a field-shuffling technique. We compute the projected distance and the velocity offsets by cross-matching the absorbers of a given quasar field with the galaxies from all the other five EIGER fields \citep{2026_Kashino}, excluding the galaxies from the original field of the absorbers from the iteration. 
Following this approach, we are able to break any real kinematic or spatial link between gas and galaxies. 
Then, we compute the covering fractions for the random sample derived from the shuffling, as we did for the original data. 
Results are shown in Fig.~\ref{dcf_500_75} for \cii\ and \civ. 
Both ions exhibit a significant covering fraction. Specifically, the \cii\ covering fraction remains above 20\% up to projected distances of 500 pkpc; at larger separations it drops to 10\% and becomes consistent with the background signal. In contrast, the \civ\ covering fraction remains significantly enhanced with respect to the background (at least of a factor of 2) up to the maximum probed separation of $\sim$ 1 pMpc.
This is due to the fact that the same \civ\ absorber can be associated with multiple galaxies at the same redshift and that 12/14 \civ\ absorption systems are associated with groups of galaxies extending to impact parameters beyond 500 pkpc (see Table~\ref{table_Carbon_absobers}).
To quantify the impact of group multiplicity, we re-calculated the \civ\ covering fraction by matching each absorber only with the galaxy with the smallest impact parameter within the considered velocity window. This restriction causes the extended plateau to completely vanish, with the covering fraction dropping to the random background level beyond R$_{\perp} \simeq 300-400$ pkpc (see Fig.~\ref{fig:closest_and_pure_civ}). This confirms that the extended highly ionized gas signal at large distances is primarily driven by group environments.

It is important to note that the covering fractions for the considered ions are not independent measurements, because there are several systems showing both \cii\ and \civ\ at the same redshift. To account for this, we computed again the covering fraction excluding from the computation of the \civ\ covering fraction the systems with associated low-ionization lines (not just \cii\ but also \oi, \siii, \feii, \alii, \mgii). The result is shown in Fig.~\ref{fig:closest_and_pure_civ}, where the enhanced covering fraction in the first bin has disappeared and it has decreased out to $\simeq 400$ pkpc; while on the larger scales the signal is still significantly higher than the background by at least of a factor two.

Then, we test the possible dependence of the covering fraction from the chosen impact parameter bin size. 
To this aim, we vary the bin edges of $\pm 5$ pkpc, and using a Monte Carlo (MC) algorithm, we investigate potential variations in the covering fraction trend, particularly for \civ\ absorbers. We performed 300 iterations, randomly shifting the bin edges within the mentioned range. From this test, we find that variations are marginal compared to the 1$\sigma$ uncertainty of the covering fraction in each interval.
This demonstrates that our results are physically driven and not affected by binning artifacts or small-number statistics.

\begin{figure*}[ht!]
    \centering
   \begin{minipage}{0.49\hsize}
       \centering
        \includegraphics[width=\linewidth]{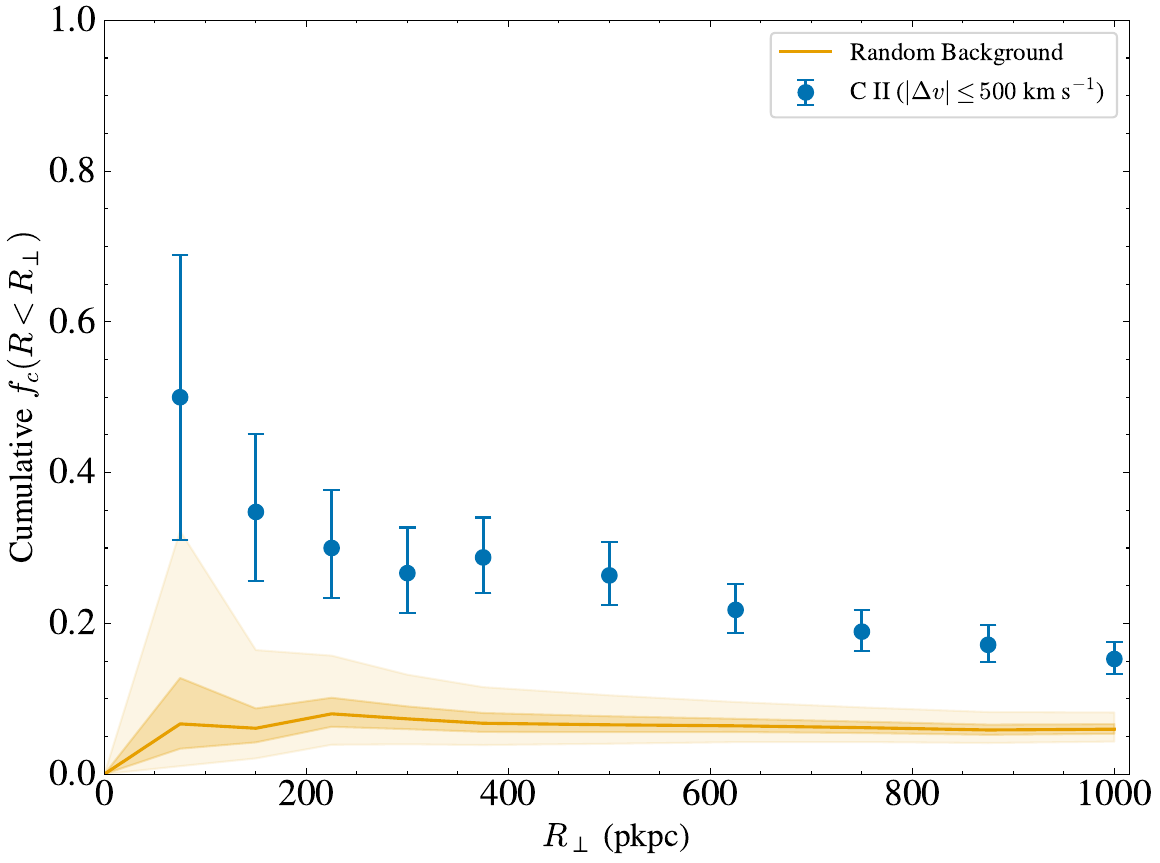}
   \end{minipage}
   \hfill
   \begin{minipage}{0.49\hsize}
       \centering
       \includegraphics[width=\linewidth]{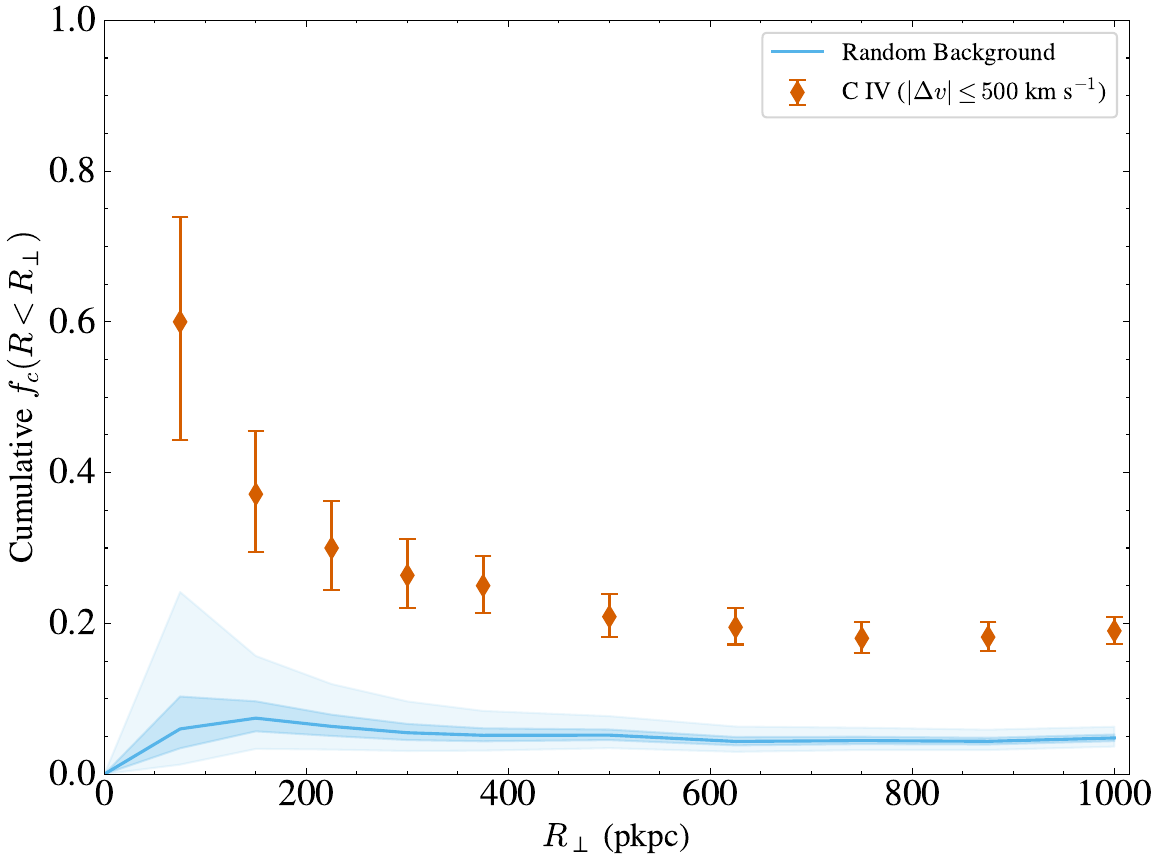}
   \end{minipage}
    \caption{Cumulative covering fraction of \cii\ (left panel) and \civ\ (right panel) absorbers as a function of the impact parameter. The points represent the measured signal within a given radius. The solid line denotes the expected random background, where the lighter and darker shaded regions indicate the its 1$\sigma$ and 3$\sigma$ confidence intervals, respectively.}
    \label{cumulative_500_75}
\end{figure*}

Finally, we compute the cumulative covering fraction. Physically, this quantity represents the average probability that a random sight-line passing anywhere within a projected distance $R_{\perp}$ from a central [\oiii] emitter intersects gas containing \cii\ or \civ. As we did for the differential covering fraction, we obtain a random background with the shuffling technique described above. In Fig.~\ref{cumulative_500_75}, we show the cumulative covering fraction, assuming $\rm \Delta v_{max} = 500$ \kms, and computing the uncertainties as Wilson-score confidence interval of 1$\sigma$.
The probability of finding a carbon absorber around an [\oiii] emitter is higher than 20\% within 600 pkpc and then drops below 20\% at distances up to maximum probe separations of $\sim$ 1 pMpc.

\subsection{Carbon absorber-Galaxy Cross-Correlation Function}
\label{sec_ccf_cg}
\begin{figure*}[ht!]
    \centering
   \begin{minipage}{0.49\hsize}
       \centering
         \includegraphics[width=\linewidth]{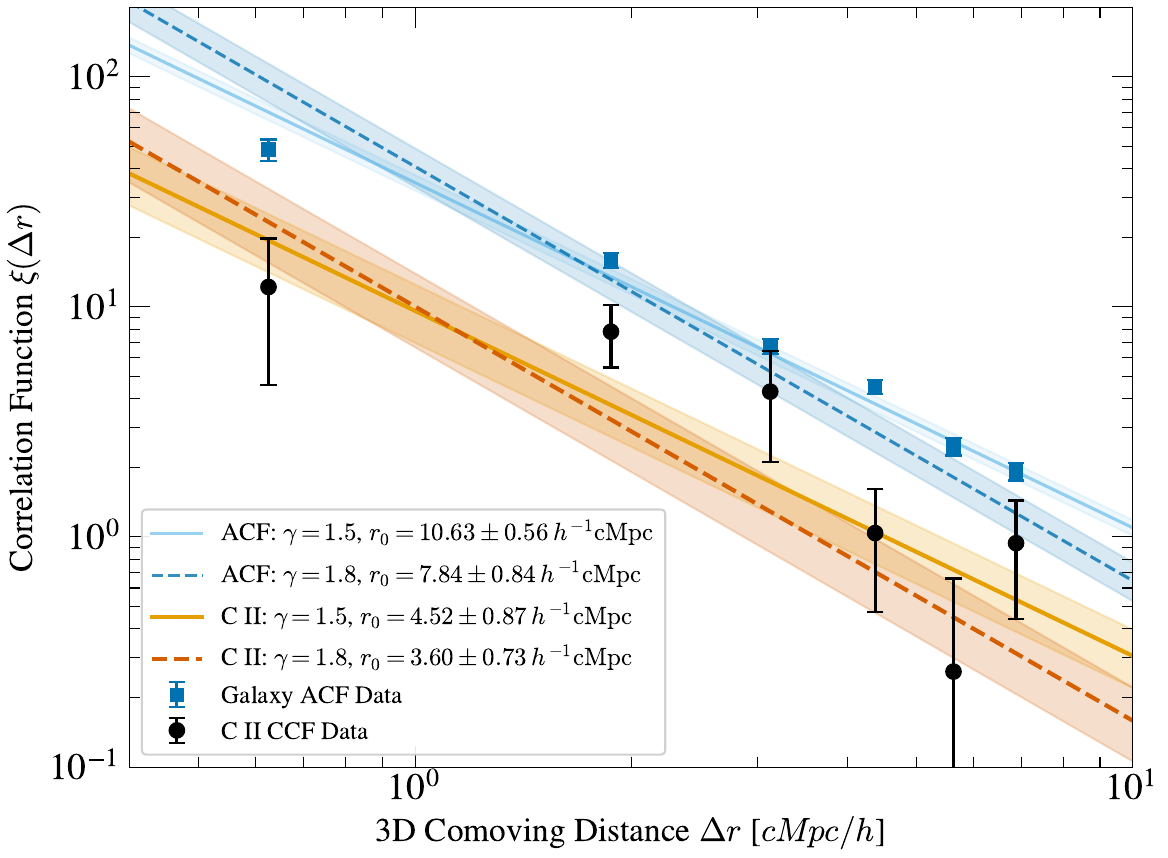}
   \end{minipage}
   \hfill
   \begin{minipage}{0.49\hsize}
       \centering
       \includegraphics[width=\linewidth]{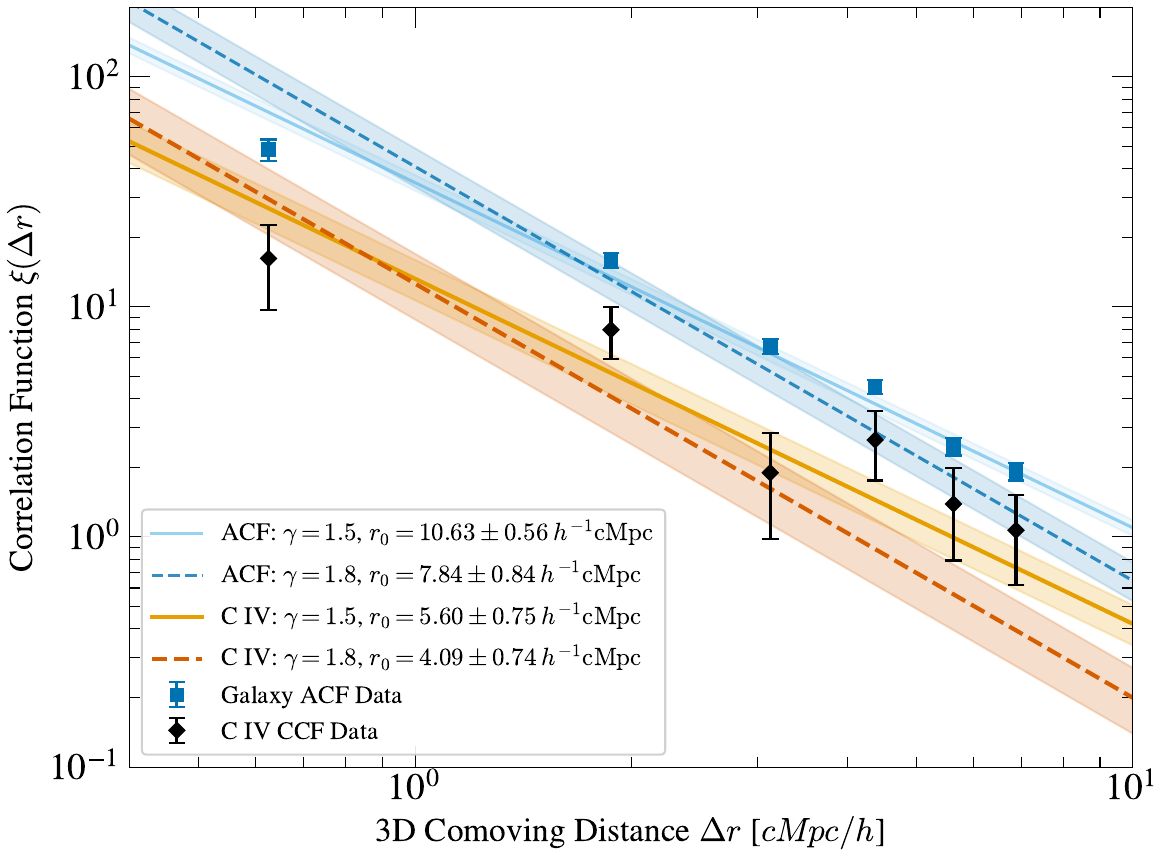}
   \end{minipage}
    \caption{3D \cii\ (left panel) and \civ\ (right panel) absorbers-[\oiii] emitting galaxy CCF as a function of the comoving distance at $z \sim 6$. The black points represent the measured signal, with vertical error bars indicating the 1$\sigma$ Poisson error. The solid orange and dotted red lines represent the best-fit power laws with index $\gamma = 1.5$ and $1.8$, respectively. The shaded orange and red regions denote the associated 1$\sigma$ confidence intervals. The blue squares denote the ACF of the [\oiii] emitters. The blue solid and dotted lines represent the best-fit power laws with index $\gamma = 1.5$ and $1.8$, respectively.} 
    \label{ccf_fit}
\end{figure*}

To capture the radial extent of the spatial clustering of the absorber-galaxy population, we evaluate the galaxy-absorber cross-correlation function (CCF). It quantifies the excess probability, relative to a random spatial distribution, of finding a \cii/\civ\ absorber at a given comoving separation from a galaxy. 
We estimate the 3D CCF, denoted as $\xi(\Delta r)$, as a function of the 3D comoving distance between a galaxy and an absorber calculated as \citep[e.g.,][]{2006_DOdorico}: 
\begin{equation}
  \rm  \Delta r = \sqrt{r_{g,\parallel}^2 + r_{a,\parallel}^2 - 2r_{g,\parallel}r_{a,\parallel} \cos(\Delta \theta)}
\end{equation}
where $\rm r_{g,\parallel}$ and $\rm r_{a,\parallel}$ are the line-of-sight comoving distances to the galaxy and the absorber, derived from their respective redshifts assuming pure Hubble flow, and $\Delta\theta$ is their angular separation on the sky. We also note that because our distances are computed in redshift space, the physical clustering signal could be affected by redshift-space distortions. In fact, the proper motions of the galaxies, combined with the peculiar velocities of the gas relative to the host halos, induce an elongation along the line of sight. Thus, to compute the correlation function, we adopt the Peebles-Hauser estimator \citep{1974ApJS...28...19P, 2013_Vargas}:
\begin{equation}
    \label{PH_estimator}
     \xi(\Delta r) = \frac{DD(\Delta r)}{DR(\Delta r)} \left( \frac{N_{DR}}{N_{DD}} \right) - 1 
\end{equation}
where $DD(\Delta r)$ represents the number of observed galaxy-absorber pairs within a given comoving distance interval $\Delta r$, while $DR(\Delta r)$ is the expected number of random pairs in the same interval. The normalization factors $N_{DD}$ and $N_{DR}$ account for the total number of real and random pairs in the entire survey volume, respectively. Regarding the $DR(\Delta r)$, we proceed using the same conceptual shuffling technique used for the covering fraction (see Sect.~\ref{covering_fraction}), constructing a random catalog that accurately reproduces the survey geometry and selection function. Specifically, to compute the $DR(\Delta r)$ pairs, we cross-match the angular coordinates of the galaxies observed in a given quasar field with the redshifts of the absorption systems detected along the sightlines of all the other EIGER fields. Finally, we compute the statistical uncertainties on the CCF considering the Poisson statistics as
\begin{equation}
    \delta\xi(\Delta r) = \frac{1 + \xi(\Delta r)}{\sqrt{DD(\Delta r)}} 
\end{equation}
We assumed comoving distance bins of 1.25 cMpc for both \cii\ and \civ\ absorbers. These intervals were chosen to correspond to $\sim180$ pkpc in proper distance, assuming a mean redshift of $z\sim6$, compatible with the CGM scale. 
We report the values of the CCF computed for both ions in Table \ref{table_CCF} and we show the corresponding results in Fig.~\ref{ccf_fit}.

\begin{table*}[h!]
\small
\caption{\cii\ and \civ\ absorbers galaxy cross-correlation function values.}               
\label{table_CCF} 
\small
\centering                        
\begin{tabular}{c c c c c c c}      
\hline\hline               
 & & \cii & & & \civ & \\
$\Delta r$ & DD($\Delta r$) & DR($\Delta r$) & $\xi(\Delta r)$ & DD($\Delta r$) & DR($\Delta r$) & $\xi(\Delta r)$ \\        
cMpc/h & & & \\
\hline                      
    0 - 1.25 & 3 & 1 & 12.20 $\pm$ 7.62 & 7 & 2 & 16.24 $\pm$ 6.52\\    
    1.25 - 2.5 & 14 & 7 & 7.80 $\pm$ 2.35 & 20 & 11 & 7.96 $\pm$ 2.00\\
    2.5 - 3.75 & 6 & 5 & 4.28 $\pm$ 2.16 & 10 & 17 & 1.90 $\pm$ 0.92\\
    3.75 - 5.0 & 13 & 28 & 1.04 $\pm$ 0.57 & 17 & 23 & 2.64 $\pm$ 0.8\\
    5.0 - 6.25 & 10 & 35 & 0.26 $\pm$ 0.40 & 16 & 33 & 1.39 $\pm$ 0.60\\
    6.25 - 7.5 & 15 & 34 & 0.94 $\pm$ 0.50 & 21 & 50 & 1.07 $\pm$ 0.45\\
\hline                                  
\end{tabular}
\tablefoot{The first column indicates the comoving distance bin, while the second and third columns represent the number of galaxy-absorber real and random pairs found in that specific bin. Finally, the last column denote the cross-correlation function.}
\end{table*}

To quantitatively describe this clustering profile and extract the characteristic correlation length, we model the measured $\rm \xi(\Delta r)$ as a power-law $\rm \xi(\Delta r) = (\Delta r/r_0 )^{-\gamma}$ where $\rm r_0$ is the correlation length and $\gamma$ is the power index. Due to the degeneracy between these two parameters, we assumed typical values for $\gamma$ adopted in the literature, $\gamma = 1.5$ and $\gamma = 1.8$ \citep{Hennawi_2006, Bielby_2011, Trainor_2012, Diener_2017, Garc9a-Vergara_2017, Garcia-Vergara_2019, 2023_Galbiati}. Thus, we obtain for the \cii\ CCF a correlation length equal to $\rm r_0 = $ 3.60 $\pm$ 0.73 cMpc/h, while for \civ\ an $\rm r_0 = $ 4.09 $\pm$ 0.74 cMpc/h, assuming $\gamma = 1.8$. As discussed in previous studies \citep[e.g.,][]{Trainor_2012, 2023_Galbiati}, the correlation length increases with decreasing slope of the power-law. Indeed, assuming $\gamma = 1.5$, the correlation lengths become $\rm r_0 = 4.52$ $\pm$ $0.87$ cMpc/h and $\rm r_0 = 5.60$ $\pm$ $0.75$ cMpc/h for \cii\ and \civ, respectively. The fact that the correlation lengths of \cii\ and \civ\ are consistent within 1$\sigma$ implies that both ionization phases share a similar large-scale clustering amplitude around high-redshift [\oiii] emitters.

\subsection{Galaxy-Galaxy Autocorrelation Function}
To quantify the spatial clustering of the population of [\oiii] emitters considered in this work, we computed the 3D galaxy-galaxy auto-correlation function (ACF), $\xi_{GG}(\Delta r)$. The ACF is computed by comparing the observed pair counts with those expected from a random distribution, subject to the exact same observational constraints. The random pairs are obtained by generating a synthetic random catalog of 500 galaxies whose angular coordinates are uniformly sampled within the same field of view of the observations. It is also important to note that we generate synthetic random galaxies without masking bright foreground objects since they only occupy small regions compared to the overall field of view. This does not affect remarkably our result due to the elevated number of galaxies generated. Furthermore, the redshifts of the random galaxies are extracted from a Gaussian Kernel Density Estimator (KDE) fitted to the redshift distribution of the full observed galaxy sample. We then calculated the comoving distance for all data-data (DD) and random-random (RR) pairs within each field with the same method used for carbon absorbers-galaxy CCF (see section \ref{sec_ccf_cg}). In this work, we assumed a comoving distance bin of 1.0 cMpc. 
We show our 3D galaxy-galaxy ACF in Fig.~\ref{ccf_fit}.

A clear and strong clustering signal is detected at a comoving distance $R < 2.5$ cMpc, reaching a significance of $\sim 13\sigma$, as expected for galaxy populations that are not randomly distributed in the Universe. It is also possible to observe a weaker, yet highly statistically significant ($\gtrsim 10 \sigma)$, extended signal up to 7.5 cMpc, probably due to galaxies peculiar velocities and large scale structures. Furthermore, following the methodology applied to the CCF, we modeled the measured galaxy-galaxy ACF as a power-law, finding the characteristic correlation lengths of $r_0^{GG} = 10.63 \pm 0.56 $ cMpc/h and $r_0^{GG} = 7.84 \pm 0.84 $ cMpc/h assuming $\gamma = 1.5$ and $\gamma = 1.8$, respectively.

For $\gamma = 1.5$, the correlation lengths obtained for ACF are higher by a factor of 2.4 and 1.9 compared to those obtained for \cii\ and \civ\ galaxy-absorbers CCFs, respectively. The two functions exhibit neither similar shapes nor the same amplitude in the EIGER fields. Thus, based on these findings, the data do not support the scenario in which [\oiii] emitters and \cii/\civ\ absorbers exclusively trace the exact same regions of the Universe. This could also suggest that a fraction of both the low- and high-ionization metals is not exclusively confined to the halos of star-forming galaxies but is instead distributed within the intergalactic medium or associated with a population of undetected faint dwarf galaxies. 

However, other previous studies have been conducted on the galaxy-galaxy autocorrelation function. Using a partially overlapping dataset of galaxies, in a redshift range of $5.95 \leq z \leq 6.55$, and in five over the six EIGER fields, \citet{2023_Eilers} reported a correlation length in the real-space of $r_0^{GG} = 4.1 \pm 0.3$ cMpc/h assuming $\gamma = 1.8$. Their analysis employs the volume-averaged projected correlation function, integrating pairs with $\Delta v < 1000$ km/s, to suppress the effects of galaxies peculiar velocities and redshift uncertainties. While projected statistics are optimal for recovering the underlying real-space clustering of galaxies, our primary objective is to perform a strict, self-consistent comparison between the galaxy-galaxy (ACF) and absorber-galaxy (CCF) clustering. By computing both the ACF and the CCF in 3D redshift space using the exact same metric and methodology, we ensure that both measurements are subject to the same line-of-sight distortions, making their relative comparison reliable.

\subsection{Equivalent Width Profile}
To explore how the absorption strength is shaped by the proximity to star-forming galaxies, we investigate the radial profile of the absorber equivalent widths. Previous literature studies found hints of an anti-correlation between the rest-frame equivalent width (REW) of \civ\ and the impact parameter of the associated galaxy. Most of these works focused on lower redshift galaxy samples: at  $z\leq1$  \citep{Chen_2001, Bordoloi_2014, Burchett_2016}, $z \leq1.5$-$2$ \citep{Dutta_2020_MAGGII}, and at $z\sim3$-$4$ \citep{2023_Galbiati}. In this work, we extend the study of the radial profile of the \civ\ absorption strength to $z\sim6$ and we compare it, for the first time, to that of the \cii\ absorbers, following the procedure described in \citep{Dutta_2020_MAGGII}. The REW is expected to decrease with increasing $R_{\perp}$ following a log-linear relationship that reproduces a sudden drop in absorption strength, scaled by a factor related to the galaxy's virial radius:
\begin{equation}
    \label{eq_ew}
    \log(W_{r}/\AA) = a + b  \cdot  (R_{\perp}/ \rm kpc)
\end{equation}
In the computation, we used both REW measurements and 3$\sigma$ upper limits determined as
\begin{equation}
    \label{equa_ew_min_col_den}
    EW_{\rm min,obs} \approx \dfrac{3 \times \lambda_{\rm ion,obs}}{R \times (S/N)_{\rm pix} \times \sqrt{n_{\rm samp}}}
\end{equation}

\noindent
where $EW_{\rm min}$ is the minimum observable equivalent width (in \AA) that can be reliably detected above the noise level, $\lambda_{obs,ion}$ is the central wavelength of the absorption feature, $R$ is the resolving power of the spectrograph, $(S/N)_{pix}$ is the median signal-to-noise ratio per pixel in the continuum of the spectrum near to the absorption feature, and finally $n_{samp}$ is the number of pixel that samples the spectral resolution element. 
Upper limits were determined at the redshift of all galaxies that did not show an associated \cii\ or \civ\ absorber within 500 pkpc from the line of sight (see Table \ref{table_Carbon_absobers}), only for the quasars for which the spectra were available. 
To compute the $(S/N)_{pix}$ a wavelength interval of  $\pm 1000$ \kms was considered centered on $\lambda_{\rm ion,obs}$, avoiding other absorption or emission features present in the spectrum. 
The resolving power for J0148, J0100, and J1030 was provided by \cite{2023_Dodorico}, while for J159 we derived the value from our analysis described in Appendix \ref{Appendix_B}. Concerning J1148 and J1120 the resolution was provided by \cite{2006_Becker} and \cite{2017_Bosman}, respectively. Finally, $n_{\rm samp}$ is computed as $n_{\rm samp} = c/(10\times R)$. 

\begin{figure*}[ht!]
    \centering
    \begin{minipage}{0.49\hsize}
       \centering
    \includegraphics[width=\linewidth]{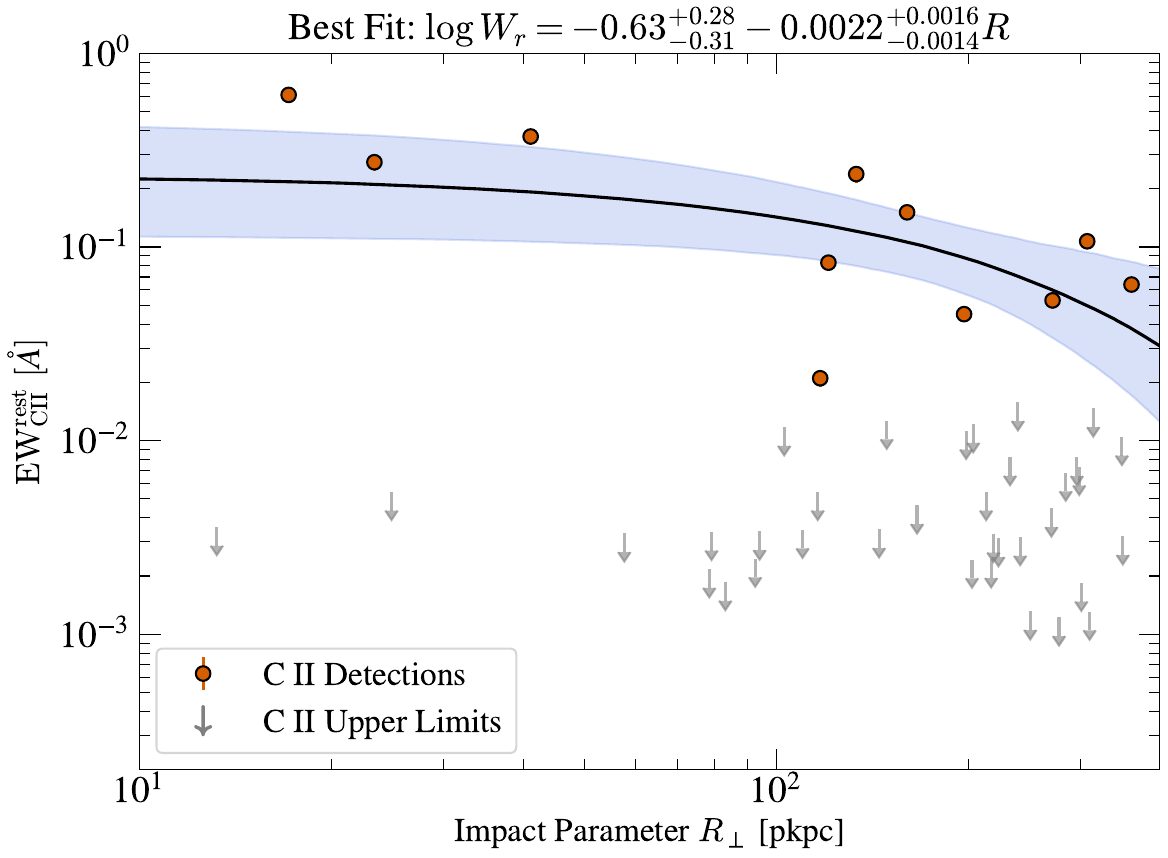}
    \end{minipage}
    \hfill
    \begin{minipage}{0.49\hsize}
       \centering
       \includegraphics[width=\linewidth]{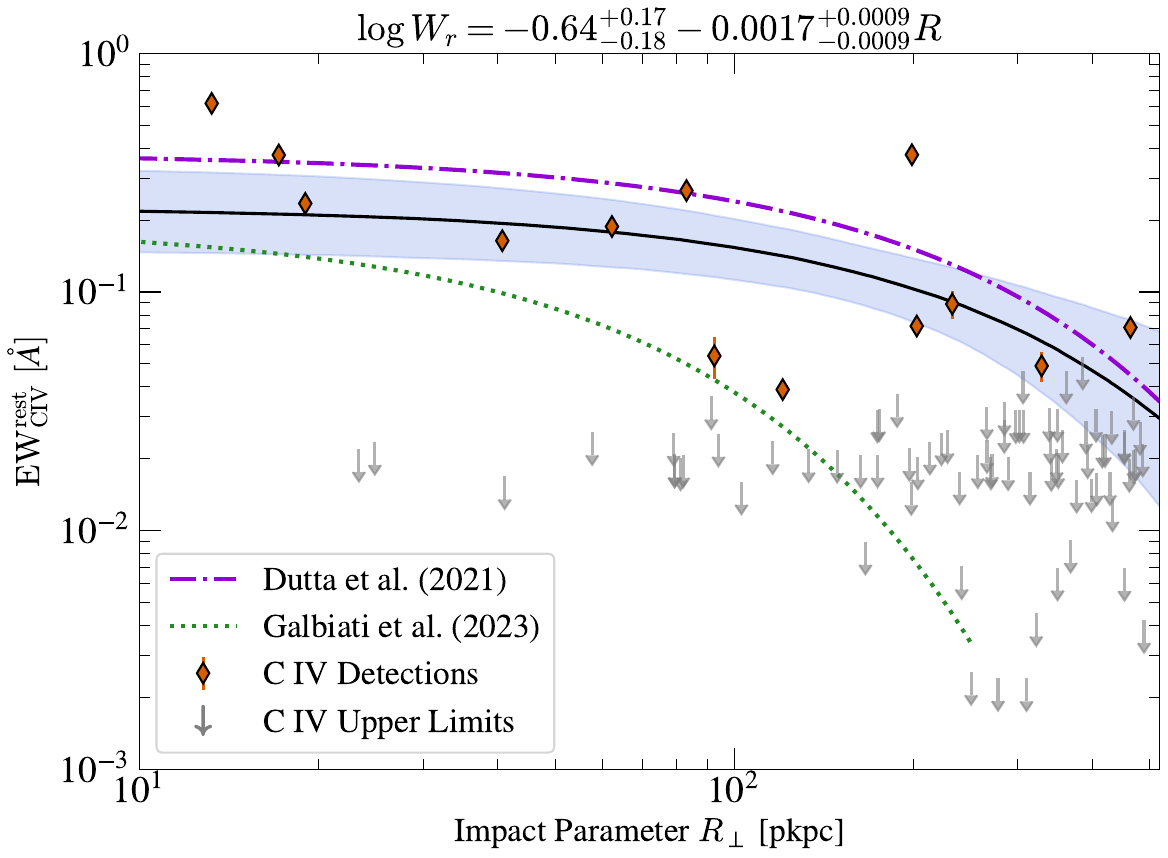}
    \end{minipage}
    \caption{EW profile as a function of transverse separation from the associated galaxy, $R_{\perp}$, for the \cii\ (left panel) and the \civ\ (right panel) absorption systems. Circles and diamonds indicate \cii\ and \civ\ detections, respectively. The 1$\sigma$ uncertainties on the rest-frame EW are typically smaller than the symbol size. Gray arrows indicate 3$\sigma$ upper limits in the EW of both carbon ions for all the galaxies that are not associated to any absorber. The best result of a log-linear fit is drawn as a solid black line and the 1$\sigma$ uncertainties are marked by shaded regions. In the right panel, we compared the results obtained for the \civ\ absorbers with the best-fit relation from \citet[][green dotted line]{2023_Galbiati} at $z=3$-$4$ and \citet[][purple dash-dotted line]{Dutta_2021} at $z<2$.} 
    \label{ew_prof}
\end{figure*}

We fit the equivalent width profile including both detections and upper limits with the function in Equation \ref{eq_ew} by applying a Bayesian method based on the product of two likelihood functions, one for measurements and one for upper limits, as described in \cite{Dutta_2020_MAGGII, Dutta_2021} \citep[for more details, see also][]{Chen_2010, Rubin_2018, 2023_Galbiati}. 
We note that in our sample there is a disparity in terms of number of detections and upper limits, especially at larger impact parameters. To take into account this disparity, we split the likelihood for the upper limits in two parts raised to two different factors, $\alpha_1$ and $\alpha_2$, respectively, considered as free parameters in the fit. The first likelihood refers to the upper limits within R$_{\perp} < 100$ pkpc and the second to those at  R$_{\perp} > 100$ pkpc. This choice is motivated by the fact that we observe only 3 detections with 10 upper limits within 100 pkpc, and 8 detection with 27 upper limits for R$_{\perp} > 100$ pkpc for \cii, while for \civ\ we have 7 detection and 10 upper limits within 100 pkpc, and 6 detections and 68 upper limits at R$_{\perp} > 100$ pkpc.
We derived posterior probability distributions and the Bayesian evidence with the nested sampling Monte Carlo algorithm MLFriends \citep{2014_Buchner, 2019_Buchner} using the UltraNest\footnote{\url{https://johannesbuchner.github.io/UltraNest/}} package \citep{2021_Buchner}.
The results are shown in Fig~\ref{ew_prof} as solid lines, while the shaded blue region marks the 1$\sigma$ confidence level. 

\section{Discussion}
\label{sec:discussion}
In this work, we have, for the first time, extended to $z>5$ the comparison between a statistical sample of \cii\ and \civ\ absorption systems and the galaxies distributed over six fields centered on $z\sim6$ quasars. 
Many previous studies investigated the same topic at lower redshift \citep{2005_Adelberger, Dutta_2021, 2021_Muzahid, 2021_Schroetter, 2023_Banerjee, 2023_Galbiati, 2024_Galbiati}, in most case identifying galaxies as Ly$\alpha$ emitters. The general conclusion of most of these works is that there is a significant enhancement of metal absorption around galaxies (extending to separations of $\sim 100-200$ pkpc) and this enhancement is more significant for galaxies in groups than for isolated ones. 
Although with our sample we do not have enough statistics to distinguish between the signal from isolated and group galaxies, we note that \civ\ absorption systems are preferentially associated with large groups of galaxies. 

\subsection{Redshift evolution of the \civ/galaxy relation}
We have investigated the signatures of redshift evolution for \civ\ absorption systems using the QSAGE sample from \citet{Dutta_2021} at $z<1$ and the MAGG sample from \citet{2023_Galbiati} at $z\sim 3$-$4$, for which we could carry out the analysis in exactly the same way as for our sample.   
However, in QSAGE galaxies are identified from their continuum emission and in some cases from the [\oii] emission, they have a median stellar mass of $\rm M_{\star} \approx 2 \times 10^9 \, M_{\odot}$, and a median SFR of $\rm 1.5 \, M_{\odot} \, yr^{-1}$ \citep{Dutta_2021}. In MAGG, instead, galaxies are Ly$\alpha$ emitters. These ranges of galaxy properties are comparable with the [\oiii] emitters found in the EIGER fields.
In order to carry out a consistent analysis, we considered in the three samples absorption systems with a \civ\ REW $\geq 0.05$ \AA. 
In Fig.~\ref{ew_prof}, we compared the REW radial profile of \civ\ absorbers in our sample with those by \citet{Dutta_2021} and \citet{2023_Galbiati}. 
The shape of the profile by \citet{Dutta_2021} is similar to our shape but with a larger normalization, while the normalization of the profile by \cite{2023_Galbiati} is in agreement with our result, but the shape differs, showing a turnaround on smaller scales. This, however, could be due to the smaller spatial extent of the MUSE field in MAGG with respect to the QSAGE and EIGER ones. 
Given the small size of our sample and the different tracers used to identify galaxies in the three samples, performing a direct comparison between these results and drawing any conclusion about a possible redshift evolution is challenging and ambiguous \citep{2023_Galbiati}.

\begin{figure}[ht!]
    \centering
    \includegraphics[width=\linewidth]{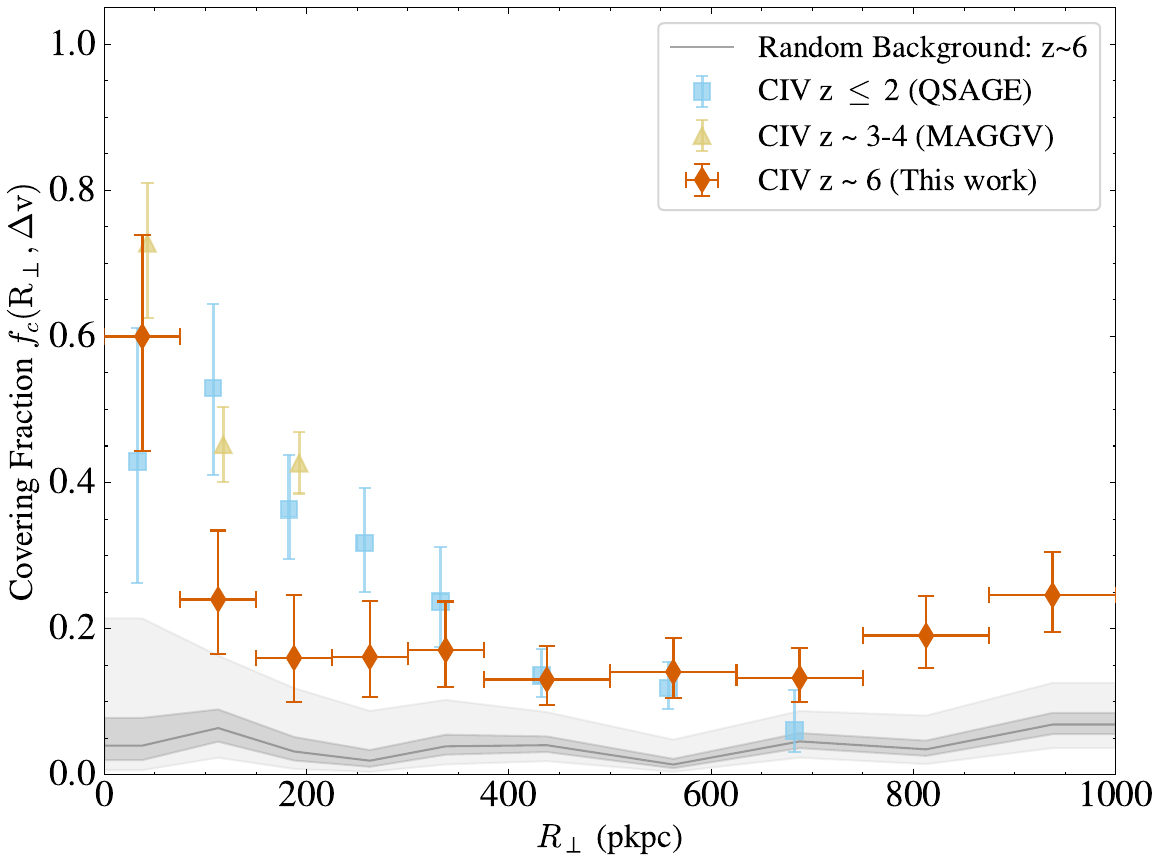}
    \caption{\civ\ covering fraction comparison between this work (orange diamonds) and lower-redshift samples: the QSAGE survey at $z \le 2$ \citep[light blue squares,][]{Dutta_2021}, and the MAGG survey at $z \sim3-4$ \citep[green triangles,][]{2023_Galbiati}. All the errors represents $1\sigma$ Wilson-score confidence interval.}
    \label{CIV_fc_comparison}
\end{figure}

\begin{figure}
    \centering
    \includegraphics[width=\linewidth]{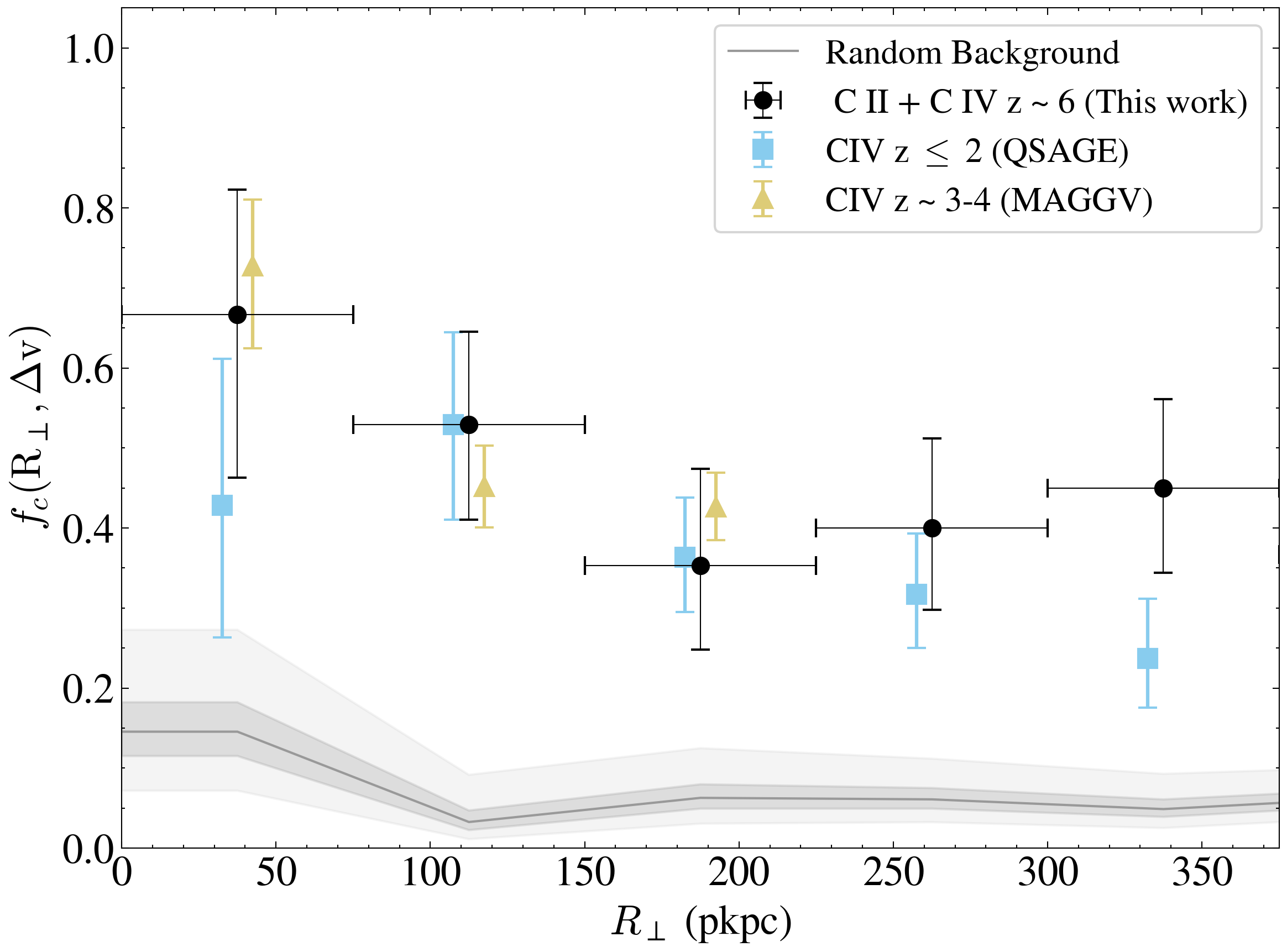}
    \caption{Same as Fig.~\ref{CIV_fc_comparison} but comparing the low-redshift \civ\ covering fractions with the \cii\ + \civ\ covering fraction from this work (black circles).  All the errors represents $1\sigma$ Wilson-score confidence interval.}
    \label{fig:cii_plus_civ_cov_frac}
\end{figure}

Figure~\ref{CIV_fc_comparison} shows the comparison of the \civ\ covering fraction for the three samples. 
In the first impact parameter bin ($R_{\perp} <75$ pkpc), the \civ\ covering fraction at $z\sim6$ is consistent, within the errors, with the values observed at lower redshift. This means that the highly ionized, metal enriched gas traced by \civ, is already present in the vicinity of galaxies at the end of Reionization. However, at larger impact parameters, the $z\sim6$ data show an evident drop compared with the lower redshift samples. This strong difference at intermediate impact parameters ($75 \leq$ R$_{\perp} \leq 400$ pkpc) might suggest a cosmic evolution in the extension of the \civ\ enriched sphere around galaxies. 
However, the interpretation of this result should take into account the rapid decline in the number density of \civ\ absorption systems observed at $z>5$ \citep[e.g.,][]{2009_RyanWeber,2011_Simcoe,2013_Dodorico,Codoreanu_2018,2022_DOdorico, davies2023b}, accompanied by an increase in the number density of low-ionization lines \citep[\oi\ and \cii,][]{Becker_2019, 2024_Sebastian}. This has been explained by the change in the ionization state of the IGM/CGM gas due to the last phases of the Reionization process.  
The observed drop in the $z\sim6$ covering fraction could be an effect of the change in the ionization state of the CGM at these redshifts rather than a difference in the distribution of the metal enriched gas. 
We tested this hypothesis by combining the covering fraction due to \civ\ and \cii, avoiding double counting if both transitions are observed at the same redshift, and restricting the redshift coverage to that of \cii\ (see Table \ref{table_covering_fraction_500}). The resulting covering fraction, shown in Fig.~\ref{fig:cii_plus_civ_cov_frac}, is now in agreement with those at lower redshift out to $\simeq 400$ pkpc. Our interpretation is that low-ionization gas traced by \cii\ significantly contribute to the CGM of $z\sim 6$ galaxies out to larger separations with respect to what happens at $z$ below the end of the Reionization process. A decisive test would be the implementation of the \cii\ $+$ \civ\ covering fraction also at lower redshift; we will explore this topic in a future work.

It is important to highlight that our measured carbon covering fractions is related to the selection function of our galaxy sample, dominated by galaxies identified via strong [\oiii] emission. This technique preferentially selects young, actively star-forming, and relatively metal-enriched systems. Since [\oiii] emitters undergo starburst-driven feedback, they are expected to launch powerful galactic winds. These outflows actively transport metal-enriched material from the interstellar medium into the halo, likely boosting the observed covering fractions of both \cii\ and \civ\ at intermediate and large impact parameters compared to a quiescent or less star-forming galaxy population.

\subsection{Carbon mass budget at $z\sim6$}
Finally, we derived a conservative lower limit for the total carbon mass associated with [\oiii]-emitting galaxies. Following the approach adopted by \cite{Bordoloi_2014} and \cite{Higginson_2026}, the mass of specific carbon ion contained in an annulus within an inner radius R$_{\rm in}$ and outer radius R$_{\rm out}$ is
\begin{equation}
     M_{\rm ion} = \dfrac{1}{f_{\rm ion}}  f_c \pi  (R_{\rm out}^2 - R_{\rm in}^2)  \Braket{N_{ion}} \ m_{\rm C}
\end{equation}
where $f_{c}$ is the covering fraction, $\Braket{N_{ion}}$ is the mean column density in the annulus, and $m_{\rm C} = 12m_p$ is the mass of carbon atom, while $m_p$ is the proton mass.\\
To ensure our estimate remains a robust lower limit, we set the ionization fraction to $f_{ion}=0.5$ for both ions. Setting $f_{ion}=0.5$ implies that all carbon resides exclusively in the \cii\ and \civ\ states, completely neglecting other ionization stages, most notably \ciii\ $\lambda\,977$ \AA\ expected to be an important fraction in diffuse CGM \citep{annurev_Tumlinson2017}. Accounting for \ciii\ is particularly challenging, as its main resonant transition falls within the Ly$\alpha$ forest, which at these redshifts becomes completely opaque due to the Gunn-Peterson effect \citep{1965_Gunn, 2006_1_Fan}.
For these reasons, the resulting mass must be strictly interpreted as the mass of the observed carbon rather than a complete mass budget. With these caveats in mind, we compute masses of each ion in two radial bins: 0-100 pkpc and 100-300 pkpc, respectively. The total mass is obtained by summing the two contributes in the two annuli considered. We find that the total \cii\ + \civ\ mass obtained, as lower limit at z$\sim$6, is $\geq 2.8 \times 10^6$ M$_{\odot}$ within R$_{\rm max}$ = 300 pkpc. This value is very similar to the total mass in oxygen obtained by \cite{Higginson_2026} of $\rm M_{O} \geq 2 \times 10^6$ M$_{\odot}$ in the same R$_{\rm max}$, considering that all oxygen is under the form of neutral oxygen, justifying this assumption due to the tight charge-exchange coupling with HI and oxygen's negligible dust depletion \citep{Bordoloi_2017, Higginson_2026}.

\section{Conclusions}
\label{sec:conclusion}
In this work, we investigated the connection between cool and ionized gas phases, traced by \cii\ and \civ\ absorption-line systems in quasar spectra, and a population of [\oiii] emitting galaxies, at $5.30<z<6.97$, identified in the EIGER survey, during the Epoch of Reionization. We traced these multiphase environments by assembling a sample of 16 \cii\ and 14 \civ\ absorbers in the redshift range $5.33 < z < 6.52$. 
Within a velocity separation of $| \Delta v | \leq 500$ km s$^{-1}$ and an impact parameter of $\rm R_{\perp} \leq 1000$ pkpc, we found 44 [\oiii] emitters around \cii\ absorbers, and 92 [\oiii] emitters from \civ\ absorbers. Leveraging these two samples, we were able to probe, for the first time, the multiphase nature of the CGM deep into the Epoch of Reionization by contrasting two distinct ionization states of carbon.
We detect an excess of both \cii\ and \civ\ absorbers around galaxies compared to random expectations, confirming that both low- and high-ionization carbon are physically associated with galaxies at $z\sim 6$. The covering fraction of both ions decreases with increasing impact parameter, although \civ\ remains detectable over larger projected distances than \cii, suggesting that the warm ionized phase is more spatially extended than the cooler gas.

The galaxy-absorber cross-correlation analysis reveals strong clustering for both ions. We measure correlation lengths of $r_0 = 3.60\pm0.73\, \rm cMpc~h^{-1}$  for \cii\ and $r_0 = 4.09\pm0.74 \, \rm cMpc~h^{-1}$ for \civ\ assuming $\gamma=1.8$, and $r_0 = 4.52\pm0.87\, \rm cMpc~h^{-1}$ for \cii\ and $r_0 = 5.60\pm0.75\, \rm cMpc\,h^{-1}$ for \civ\ assuming $\gamma=1.5$. However, the galaxy-absorber correlation remains significantly weaker than the [\oiii] emitters auto-correlation function, which has a correlation length of $r_0^{GG} = 7.79 \pm 0.64 \, \rm cMpc\, h^{-1}$ assuming $\gamma = 1.8$, and $r_0^{GG} = 9.12 \pm 0.85 \, \rm cMpc \, h^{-1}$ assuming $\gamma = 1.5$. This suggests that a fraction of both low- and high-ionization carbon is not exclusively confined to the halos of massive galaxies, but is likely distributed throughout the surrounding large-scale environment or associated with galaxies fainter than the detection limit of the EIGER survey (40\% complete at $L_{5008} = 10^{42}$ erg s$^{-1}$).

The equivalent width profiles of both ions show a decline with increasing impact parameter. Focusing on \civ, it seems to be consistent with previous measurements at lower redshift, such as QSAGE at $z \leq 2$ and MAGG at $z\sim 3-4$. In particular, the \civ\ profile agrees with measurements from the MAGG survey until 100 pkpc, suggesting that the radial distribution of highly ionized carbon around galaxies may already be largely established by the end of the EoR.

In addition, we derive a conservative lower limit on the observed carbon mass associated with galaxies of $M_{\cii + \civ}\simeq 2.8\times10^6\,{\rm M_\odot}$ within 300 pkpc. Since our estimate neglects unobserved ionization states, particularly \ciii, it should be interpreted as a lower limit to the observed carbon reservoir rather than the total carbon budget of the CGM.

Finally, by comparing our measurements with lower-redshift surveys, we find that the \civ\ covering fraction at small impact parameters is already comparable to that observed at later cosmic times, while it declines more rapidly at intermediate distances. This behavior is consistent with the observed evolution of the cosmic abundance of carbon ions and supports a scenario in which a larger fraction of circumgalactic carbon resides in lower ionization states during the Epoch of Reionization. 

Overall, our results provide one of the first statistical characterizations of the multiphase CGM around star-forming galaxies during the Epoch of Reionization using two ionization states of the same element. They indicate that chemically enriched gas was already widespread around galaxies at $z\sim6$, while simultaneously suggesting that the ionization structure and spatial distribution of circumgalactic metals were still evolving. In a future paper (Galbiati et al. in prep.), we will investigate the connection between absorbers and galaxies identified through Ly$\alpha$ emission, over more than 30 fields of quasars at $z\sim6$.  It would also be desirable to increase the number of quasar fields studied with JWST with the same approach adopted in the EIGER survey. All these studies would enable a more complete picture of the multiphase CGM and its role in the early assembly of galaxies. 

\begin{acknowledgements}
      We thank the referee for the careful reading of this paper and the suggestions which improved the manuscript.
      CP, VD, KK, LPe, LPa, and RR acknowledge support from the ERC synergy grant 101166930 - RECAP.
      FS acknowledges financial support from the Bando Finanziamento ASI CI-UCO-DSR-2022-43 CUP:C93C25004260005 project ''IBISCO: feedback and obscuration in local AGN'', the Ricerca Fondamentale INAF 2023 Data Analysis grant 1.05.23.03.04 ''ARCHIE ARchive Cosmic HI \& ISM  Evolution'', Ricerca Fondamentale INAF 2024 under project 1.05.24.07.01 MINI-GRANTS RSN1 "ECHOS".
      CM acknowledges support from Fondecyt Iniciacion grant 11240336 and the ANID BASAL project FB210003.
      This work is based on observations collected at the European Southern Observatory under ESO programmes 098.B-0537(A) and 0114.B-0278(A).
      This research made use of \texttt{Astropy} \citep{2013_Astropy, 2018_Astropy, 2022_Astropy}, \texttt{scipy} \citep{2020SciPy-NMeth}, \texttt{numpy} \citep{harris2020array}, \texttt{lmfit} \citep{2021_Newville}, \texttt{matplotlib} \citep{Hunter_2007}, and \texttt{pandas} \citep{mckinney-proc-scipy-2010}.
\end{acknowledgements}
\bibliographystyle{bibtex/aa}
\bibliography{bibliography}

\begin{appendix}
\nolinenumbers
\onecolumn
\section{Carbon absorbers with associated [\oiii] emitters}
\label{Appendix_A}
\begingroup
\tiny
\begin{longtable}{c c c c c c c c c}
\caption{\cii\ and \civ\ absorption systems with $\log (N) \ + \sigma_{\log(N)} \geq 13.0 $, and their associated galaxies in the EIGER Survey fields. For each absorption system, we list all the associated galaxies found within $R_{\perp} \leq 1000 \: \mathrm{kpc}$ and $ |\Delta v | < 500 $ \kms.} 
\label{table_Carbon_absobers} \\
\hline\hline 
Quasar & $z_{\mathrm{abs}}$ & log(N$_{\cii}$) & log(N$_{\civ}$) & Gal ID & $z_{gal}$ & $R_{\perp}$ & $\Delta v$ & $\Delta \rm r$\\ 
Field & & cm$^{-2}$ & cm$^{-2}$ & & & pkpc & \kms & cMpc \\
\hline
\endfirsthead
\endfoot
\hline
\endlastfoot
   J0148 & 5.48774 & <12.09 & $13.17 \pm 0.13$ & 19570 & 5.490 & 92.6 & 104.4 & 1.27\\
    & & <12.09 & & 19764 & 5.490 & 292.1 & 104.4 & 2.20\\
   J1030 & 5.51737 & - & $13.74 \pm 0.07$ & 7771 & 5.519 & 40.7 & -75.0 & 0.85\\
    & & & & 7594 & 5.516 & 46.6 & -63.0 & 0.74\\
    & & & & 16469 & 5.518 & 88.3 & 29.0 & 0.65\\
    & & & & 8631 & 5.517 & 115.9 & -17.0 & 0.78\\
    & & & & 12984 & 5.516 & 362.9 & -247.1 & 3.55\\
    & & & & 6206 & 5.509 & 866.8 & -385.3 & 6.99\\
    & 5.72501 & <12.25 & $14.56 \pm 0.04$ & 7377 & 5.728 & 13.2 & 133.0 & 1.41\\
    & & <12.24 & & 7150 & 5.721 & 109.9 & -178.8 & 2.03\\
    & & <12.24 & & 7319 & 5.722 & 144.9 & -134.2 & 1.72\\
    & & <12.23 & & 6714 & 5.724 & 219.1 & -45.0 & 1.55\\
    & & <12.23 & & 12196 & 5.720 & 452.2 & -223.4 & 3.85 \\
    & & & & 15831 & 5.726 & 609.3 & 44.1 & 4.12 \\
    & & & & 13611 & 5.727 & 613.8 & 88.7 & 4.23 \\
    & & & & 16565 & 5.733 & 632.6 & 356.0 & 5.68 \\
    & & & & 12674 & 5.719 & 958.5 & -268.0 & 7.04 \\
    & 5.74254 & $14.69 \pm 0.01$ & $14.08 \pm 0.03$ & 6949 & 5.746 & 17.1 & 153.8 & 1.63\\ 
    & & & & 15869 & 5.746 & 174.3 & 153.8 & 2.00 \\
    & & & & 3116 & 5.747 & 488.0 & 198.2 & 3.75 \\
    & & & & 2766 & 5.748 & 493.7 & 242.7 & 4.03 \\
    & & & & 5740 & 5.739 & 519.6 & -157.4 & 4.01 \\
    & & & & 16565 & 5.733 & 632.6 & -424.5 & 6.40 \\
    & 5.96731 & <12.61 & $13.58 \pm 0.28$ & 8512 & 5.969 & 232.6 & 72.7 & 1.79\\ 
    & & <12.61 & & 8941 & 5.968 & 296.3 & 29.7 & 2.09\\
    & & <12.56 & & 9834 & 5.959 & 298.8 & -357.8 & 4.26 \\
    & & <12.61 & & 9879 & 5.969 & 336.0 & 72.7 & 2.46\\
    & & <12.54 & & 8588 & 5.956 & 383.7 & -487.0 & 5.72 \\
    & & & & 30001 & 5.963 & 579.8 & -185.5 & 4.47 \\
    & & & & 8048 & 5.957 & 740.0 & -444.0 & 6.91 \\
    & & & & 14036 & 5.962 & 760.8 & -228.1 & 5.80 \\
    & & & & 8861 & 5.961 & 769.1 & -271.6 & 6.05 \\
    & & & & 5862 & 5.961 & 774.5 & -271.6 & 6.08 \\
    & & & & 8251 & 5.961 & 930.2 & -271.6 & 7.06 \\
    & & & & 12296 & 5.958 & 968.2 & -400.9 & 7.92\\
   J0100 & 5.33848 & - & $13.92 \pm 0.01$ & 9193 & 5.338 & 19.0 & -22.7 & 0.27\\
    & & & & 5271 & 5.342 & 373.9 & 166.4 & 2.98 \\
    & & & & 16588 & 5.336 & 717.8 & -177.3 & 4.72 \\ 
    & 5.51325 & - & $13.10 \pm 0.10$ & 5581 & 5.510 & 328.9 & -149.6 & 2.68 \\
    & & & & 5255 & 5.512 & 420.9 & -57.5 & 2.81 \\
    & & & & 1659 & 5.508 & 497.8 & -241.7 & 4.15 \\
    & & & & 16252 & 5.514 & 506.2 & 34.5 & 3.32 \\
    & & & & 13791 & 5.510 & 736.4 & -149.6 & 5.06 \\
    & & & & 17125 & 5.523 & 745.4 & 448.4 & 6.84 \\
    & & & & 11987 & 5.505 & 784.1 & -380.0 & 6.53 \\
    & 5.79746 & $14.10 \pm 0.01$ & $12.98 \pm 0.02$ & - & - & - & - & - \\
    & 5.94474 & $13.79 \pm 0.06$ & $12.99 \pm 0.04$ & 7406 & 5.943 & 120.6 & -75.1 & 1.14\\ 
    & & & & 5394 & 5.941 & 215.8 & -161.5 & 2.25 \\
    & & & & 6006 & 5.941 & 284.1 & -161.5 & 2.59 \\
    & & & & 4080 & 5.942 & 309.3 & -118.3 & 2.47 \\
    & & & & 4032 & 5.943 & 318.0 & -75.1 & 2.34 \\
    & & & & 30 & 5.944 & 586.1 & -31.9 & 4.08 \\
    & 6.01122 & <12.08 & $13.31 \pm 0.13$ & 6172 & 6.014 & 202.8 & 118.8 & 1.88 \\
    & & & & 14517 & 6.000 & 610.4 & -480.1 & 6.55 \\
    & & & & 2159 & 6.008 & 759.9 & -137.7 & 5.51 \\
    & 6.11162 & $14.09 \pm 0.08$ & - & - & - & - & - & - \\
    & 6.14349 & $14.10 \pm 0.01$ & - & - & - & - & - & - \\
    & 6.18667 & <11.97 & $13.95 \pm 0.02$ & 7216 & 6.188 & 83.1 & 55.9 & 0.83 \\
    & & <11.96 & & 4184 & 6.187 & 301.1 & 14.2 & 2.17 \\
    & & <11.96 & & 9273 & 6.187 & 487.1 & 14.2 & 3.50 \\
    & & & & 8514 & 6.182 & 511.4 & -194.5 & 4.18 \\
    & & & & 17352 & 6.184 & 544.2 & -111.0 & 4.07 \\
    & & & & 30001 & 6.186 & 552.6 & -27.5 & 3.98 \\
    & & & & 2355 & 6.182 & 644.9 & -194.5 & 5.04 \\
    & & & & 3783 & 6.183 & 650.7 & -152.7 & 4.93 \\
    & & & & 9270 & 6.184 & 658.0 & -111.0 & 4.86 \\
    & & & & 8651 & 6.182 & 698.0 & -194.5 & 5.39 \\
    & & & & 421 & 6.193 & 702.3 & 264.4 & 5.73 \\
    & & & & 8072 & 6.192 & 760.6 & 222.7 & 5.92 \\
    & & & & 13753 & 6.180 & 790.5 & -278.0 & 6.35 \\
    & & & & 15722 & 6.183 & 816.3 & -152.7 & 6.07 \\
    & & & & 9148 & 6.178 & 822.1 & -361.5 & 6.96 \\
    & & & & 18057 & 6.193 & 894.1 & 264.4 & 6.97 \\
    & & & & 8343 & 6.177 & 905.9 & -403.2 & 7.70 \\
    & & & & 13073 & 6.175 & 909.0 & -486.8 & 8.21 \\
    & & & & 6083 & 6.185 & 921.0 & -69.3 & 6.66 \\
    & & & & 11506 & 6.186 & 921.7 & -27.5 & 6.63 \\
    & & & & 13590 & 6.188 & 922.9 & 55.9 & 6.66 \\
    & & & & 6365 & 6.180 & 985.3 & -278.0 & 7.62 \\
    & & & & 11400 & 6.182 & 989.1 & -194.5 & 7.38 \\
    & & & & 15162 & 6.193 & 993.6 & 264.4 & 7.64 \\
   J159 & 5.73484 $\pm$ 0.00001 & $14.671 \pm 0.012$ & - & - & - & - & - & - \\ 
    & 5.91269 $\pm$ 0.00002 & $14.415 \pm 0.001$ & <12.735 & 9549 & 5.906 & 133.4 & -290.4 & 3.16 \\
    & & & <12.741 & 15362 & 5.913 & 197.0 & 13.4 & 1.37 \\
    & & & <12.741 & 15408 & 5.913 & 358.8 & 13.4 & 2.48\\
    & & & <12.739 & 12178 & 5.917 & 472.8 & 186.8 & 3.80 \\
    & 5.92243 $\pm$ 0.00002 & $13.507 \pm 0.041$ & - & 12305 & 5.920 & 638.9 & -105.3 & 4.56 \\
    & 6.0554 $\pm$ 0.0001 & $14.267 \pm 0.039$ & <12.622 & 7921 & 6.051 & 41.1 & -199.3 & 2.08 \\
    & 6.21904 $\pm$ 0.0001 & $13.423 \pm 0.084$ & <12.715 & 8654 & 6.219 & 271.3 & -1.8 & 1.96 \\
    & & & & 7617 & 6.208 & 555.2 & -458.9 & 6.16 \\
    & & & & 17520 & 6.213 & 631.0 & -251.1 & 5.22 \\
    & & & & 8316 & 6.211 & 821.8 & -334.2 & 6.84 \\
    & 6.2385 $\pm$ 0.0001 & $14.134 \pm 0.024$ & <12.737 & 15962 & 6.239 & 23.4 & 18.4 & 0.25 \\
    & & & <12.714 & 9215 & 6.241 & 79.3 & 101.2 & 1.18 \\
    & & & <12.701 & 17057 & 6.244 & 81.2 & 225.4 & 2.37 \\
    & & & <12.774 & 16185 & 6.236 & 266.2 & -105.9 & 2.21 \\
    & & & <12.701 & 16662 & 6.245 & 288.9 & 266.8 & 3.43 \\
    & & & <12.794 & 9580 & 6.241  & 429.1 & 101.2 & 3.27 \\
    & & & <12.759 & 16421 & 6.248 & 486.7 & 390.9 & 5.32 \\
    & & & & 12958 & 6.247 & 841.9 & 349.5 & 7.06 \\
   J1148$^a$ & 6.00964 $\pm$ 0.00005 & $14.171 \pm 0.065$ & - & 18887 & 6.011 & 160.3 & 55.6 & 1.26 \\ 
    & & & & 7081 & 6.015 & 305.6 & 226.6 & 3.18 \\ 
    & & & & 10731 & 6.009 & 315.5 & -29.9 & 2.23 \\ 
    & 6.12912 $\pm$ 0.00007 & $13.958 \pm 0.055$ & - & 8739 & 6.128 & 307.5 & -54.7 & 2.26 \\
    & 6.19678 $\pm$ 0.00001 & $13.027 \pm 0.072$ & - & 11725 & 6.206 & 117.1 & 383.0 & 4.00 \\
    & 6.25542 $\pm$ 0.00005 & $13.731 \pm 0.104$ & - & 9422 & 6.257 & 361.0 & 62.0 & 2.69 \\
    & & & & 2852 & 6.253 & 555.8 & -103.3 & 4.17 \\
    & & & & 8234 & 6.256 & 740.2 & 20.7 & 5.37 \\
    & & & & 21959 & 6.250 & 864.4 & 227.3 & 6.68 \\
    & & & & 17433 & 6.256 & 935.0 & 20.7 & 6.79 \\
   J1120$^a$ & 5.79539 & - & $13.97 \pm 0.03$ & 14866 & 5.792 & 199.0 & -149.6 & 2.07 \\ 
    & & & & 7955 & 5.795 & 243.0 & -17.2 & 1.66 \\ 
    & 6.17111 & - & $13.67 \pm 0.03$ & 8334 & 6.168 & 62.3 & -129.6 & 1.40 \\
    & & & & 6929 & 6.171 & 356.5 & -4.2 & 2.56 \\
    & & & & 7013 & 6.169 & 374.4 & -87.8 & 2.83 \\
    & & & & 8253 & 6.171 & 714.5 & -4.2 & 5.12 \\
    & 6.40671 & $13.4 \pm 0.4$ & - & - & - & - & - & - \\ 
    & 6.51111 & - & $13.25 \pm 0.06$ & 11915 & 6.516 & 463.9 & 35.5 & 3.50 \\
    & & & & 11636 & 6.511 & 485.5 & -164.0 & 4.00 \\
    & & & & 11300 & 6.514 & 578.9 & -44.9 & 4.37 \\
    & & & & 10566 & 6.516 & 768.1 & 35.5 & 5.78 \\
    & & & & 1526 & 6.516 & 820.2 & 35.5 & 6.17 \\
    & & & & 12862 & 6.505 & 894.5 & -403.6 & 7.83 \\
    & & & & 11713 & 6.525 & 920.8 & 394.3 & 7.97 \\
    & & & & 11618 & 6.526 & 946.1 & 434.1 & 8.34\\
\end{longtable}
\tablefoot{The first column denote the quasar field. The second the redshift of the absorption system. The third and fourth denote the column density of \cii\ and \civ\ absorption system. The fifth and sixth column denote the galaxy ID and its redshift provided by \cite{2026_Kashino}. The seventh column indicate the impact parameter of each galaxy. The last two column describe the line-of-sight distance for every absorber-galaxy pair as a velocity offset and the 3D comoving distance between a galaxy and an absorber. \\
We report upper limits only for galaxies with R$_{\perp} \leq 500$ pkpc, used for the REW radial profile analysis.\\
$^a$ For J1148 and J1120 we do not report \civ\ upper limits due to the unavailable spectrum of the quasar.}
\twocolumn
\begin{figure}[ht!]
    \centering
    \includegraphics[width=\linewidth]{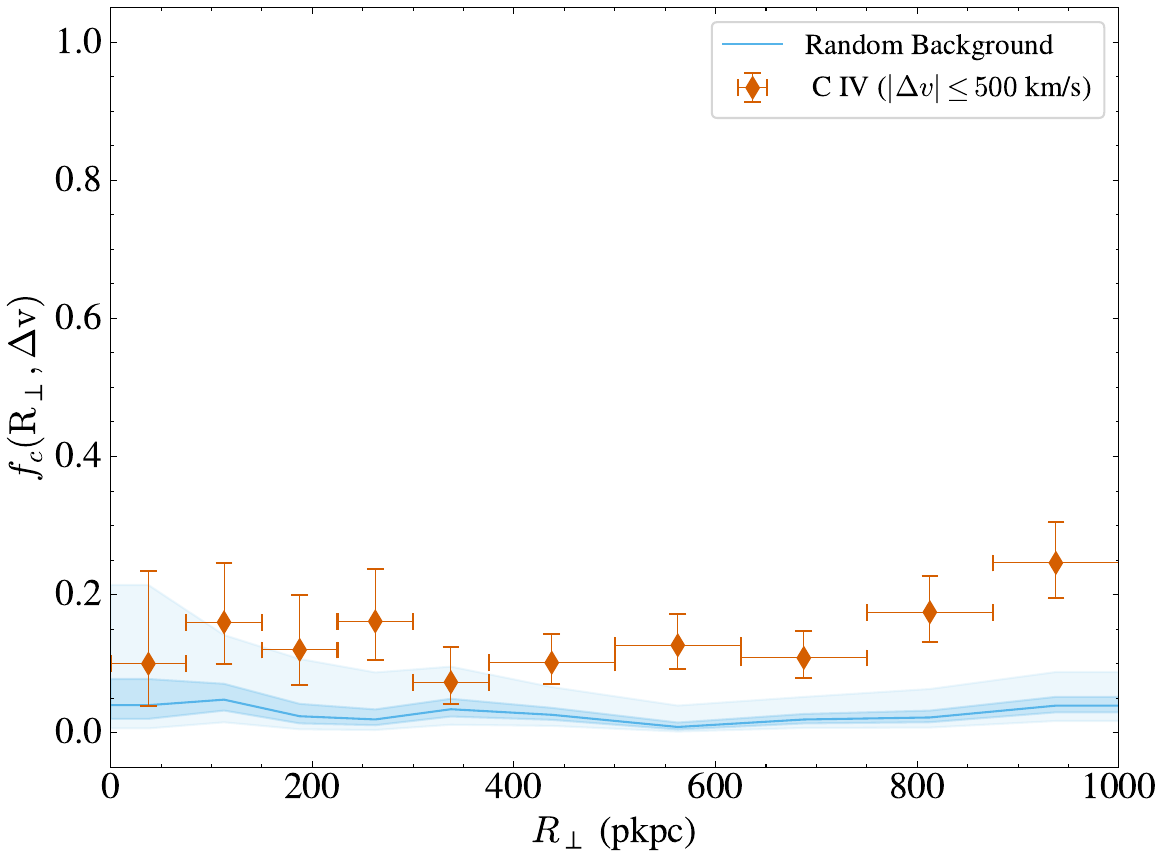}
    \includegraphics[width=\linewidth]{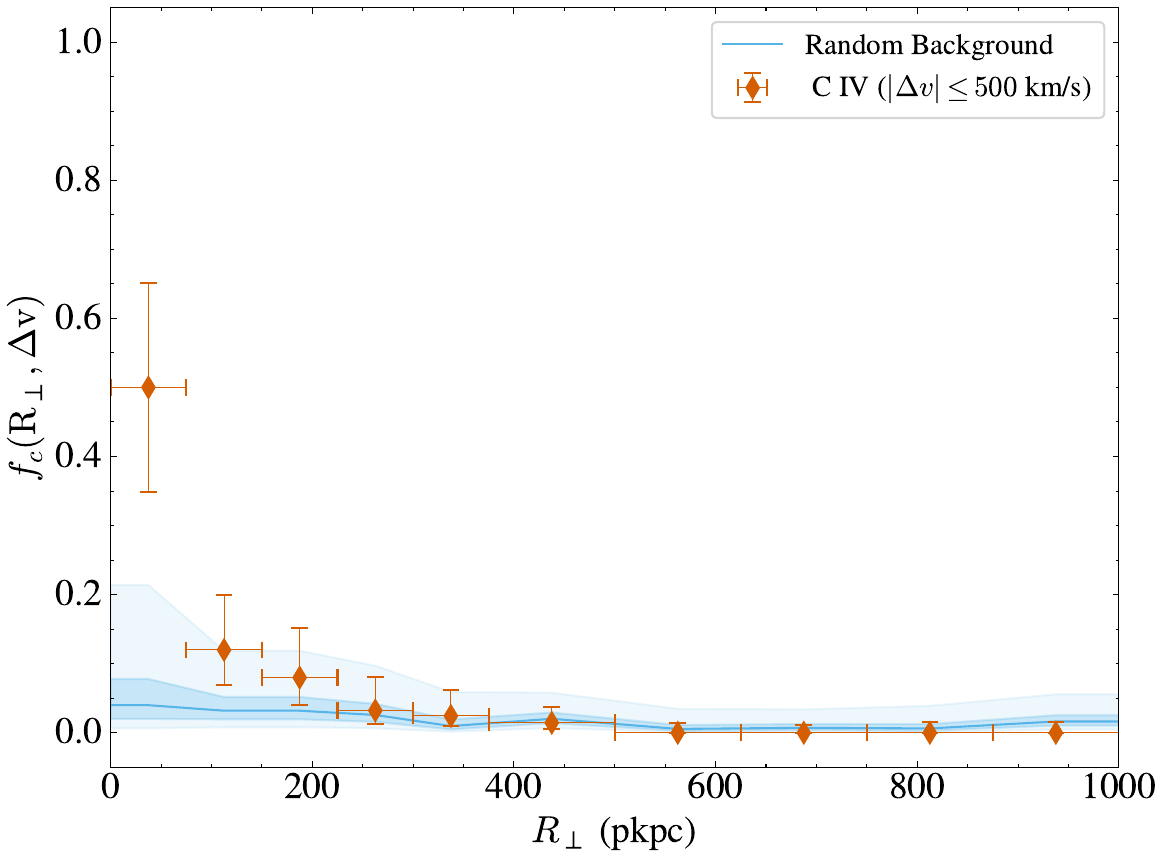}
    \caption{The upper panel represents the \civ\ covering fraction obtained excluding the systems that are also associated with low-ionization absorption lines. The lower panel represents the \civ\ covering fraction computed considering only the association between \civ\ absorbers and the closest galaxy.}
    \label{fig:closest_and_pure_civ}
\end{figure}
\endgroup
\section{The new X-Shooter spectrum of PSOJ159-02}
\label{Appendix_B}
In this appendix, we present the new VLT/X-Shooter observations of PSO J159-02 (0114.B-0278; PI: L. Zappacosta), which significantly improve the S/N of previous available observations (098.B-0537; PI: E. Farina). Table \ref{Table_XShooter_Obs} summarizes the observational details of the observing program.
\begin{table*}
\caption{Summary of VLT/X-Shooter observations for PSO J159-02.}
\label{Table_XShooter_Obs} 
\tiny
\centering                        
\begin{tabular}{c c c c c}      
\hline\hline               
X-Shooter Proposal ID & P.I. & Observation date & Exposure time & \# of observation\\
 &  &  & (s) & \\
\hline
098.B-0537(A) & E.P. Farina & 04 Apr 2017 & 1200 & 2\\
 & & 17 Dec 2017 & 1200 & 2\\
0114.B-0278(A) & L. Zappacosta & 25 Jan 2025 & 900 & 8\\
 &  & 28 Jan 2025 & 900 & 4\\
 &  & 29 Jan 2025 & 900 & 4\\
 &  & 04 Feb 2025 & 900 & 4\\
 &  & 05 Feb 2025 & 900 & 4\\
\hline                      
\end{tabular}
\tablefoot{Observation log detailing the two X-Shooter programs utilized in this work.}
\end{table*}
All the data were collected with a slit-width of $0.9''$ for the VIS arm and $0.6''$ for the NIR one. 
Concerning the resolution, we obtain $R = \lambda/\Delta \lambda$ of 10700 and 8100 for VIS and NIR, respectively. In particular, the NIR resolution is the nominal resolution of the arm, due to the fact that the seeing at the time of observations is larger than the slit. On the other hand, for the VIS arm, being the seeing lower than the slit, we follow the discussion in \cite{2023_Dodorico} sec. 5.2, to determine the effective resolving power, adopting
\begin{equation}
    \rm R_{eff} = \frac{R_{nom}\times slit}{FWHM_{seeing}}
\end{equation}
and achieving a resolution of 10700. 
The data reduction has been performed using the open-source python software \texttt{PypeIt} \citep{2019_Pypeit, 2020_Pypeit}, following the standard procedures for UV and NIR spectroscopic data \citep{2020_S, 2022_Farina}.\\
As a first step we corrected each single exposures for all the instrumental signatures (e.g., flat-field, bias). 
Wavelength calibration was performed using Th-Ar arc lamp exposures for the VIS arm, while for the NIR arm, the wavelength solution was derived utilizing OH sky emission lines \citep{2000_Rousselot} to ensure higher accuracy and account for instrumental flexures.
Then, cosmic rays were removed through the \texttt{L.A. Cosmic} algorithm \citep{2001_VDK}.
For the NIR spectrum to perform sky subtraction on the 2D images difference imaging of dithered AB pairs and a 2D BSpline fitting procedure were adopted.
The quasar trace was identified automatically and then extracted through optimal extraction \citep{1986_Horne} in order to produce a 1-D spectrum.  Each 1-D spectrum was flux-calibrated using X-Shooter standard stars. All the 1-D VIS flux-calibrated spectra were then co-added on a new velocity grid, choosing a step of 10 km/s. We treated the NIR spectra similarly.
As a last step, telluric absorption features were removed from the co-added and stacked spectrum using telluric grids generated by the Line-by-Line Radiative Transfer Model (\texttt{LBLRTM}) \citep{CLOUGH_2005, Gullikson_TelFit}.

The two final spectra for the VIS and NIR arm were then stitched together following the procedure described in \cite{2023_Dodorico}. We scaled the NIR spectrum to the VIS one computing, at first, the median value in the overlap region 1000-1020 $\mu$m for both arms, and then estimating the ratio between these two median. After multiplying the NIR spectrum by this scale factor, we cut the VIS reddest and NIR bluest part in order to avoid the noisiest region at 1017 nm.
Finally, we performed the absolute flux-calibration scaling the final spectrum to the J-band photometry of the object, J = 20.00 $\pm$ 0.10, as reported in \cite{Banados_2016}. 

\subsection{Fit of the continuum and rest-UV properties}
To characterize the rest-frame UV-optical properties of our quasar, we used the typical procedure for fitting the spectrum of high-z quasars largely discussed in the literature \citep[e.g.,][]{Mazzucchelli_2017, 2020_S, 2022_Farina, Mazzucchelli_2023, 2023_Bischetti}. 
The rest-frame UV/optical spectrum of quasars is dominated by radiation from the accretion disk, which is well modeled by a power-law component. Additionally, blended high-order Balmer lines and bound-free Balmer continuum emission give rise to a Balmer pseudo-continuum, which is a non-negligible component in the considered wavelength range.
Finally, there is the so-called iron pseudo-continuum. This component is mainly due to transitions of single- and double-ionized iron atoms (\feii\ and \feiii), and it is particularly strong around the broad \mgii\ emission lines.
The complete continuum model is fitted to regions that are considered free from narrow and broad emission lines, but also of absorption and telluric lines \citep{2010_Decarli,Shen_2011,Mazzucchelli_2017, 2020_S}.

\begin{figure*}[ht!] 
    \centering
    \includegraphics[width=\textwidth]{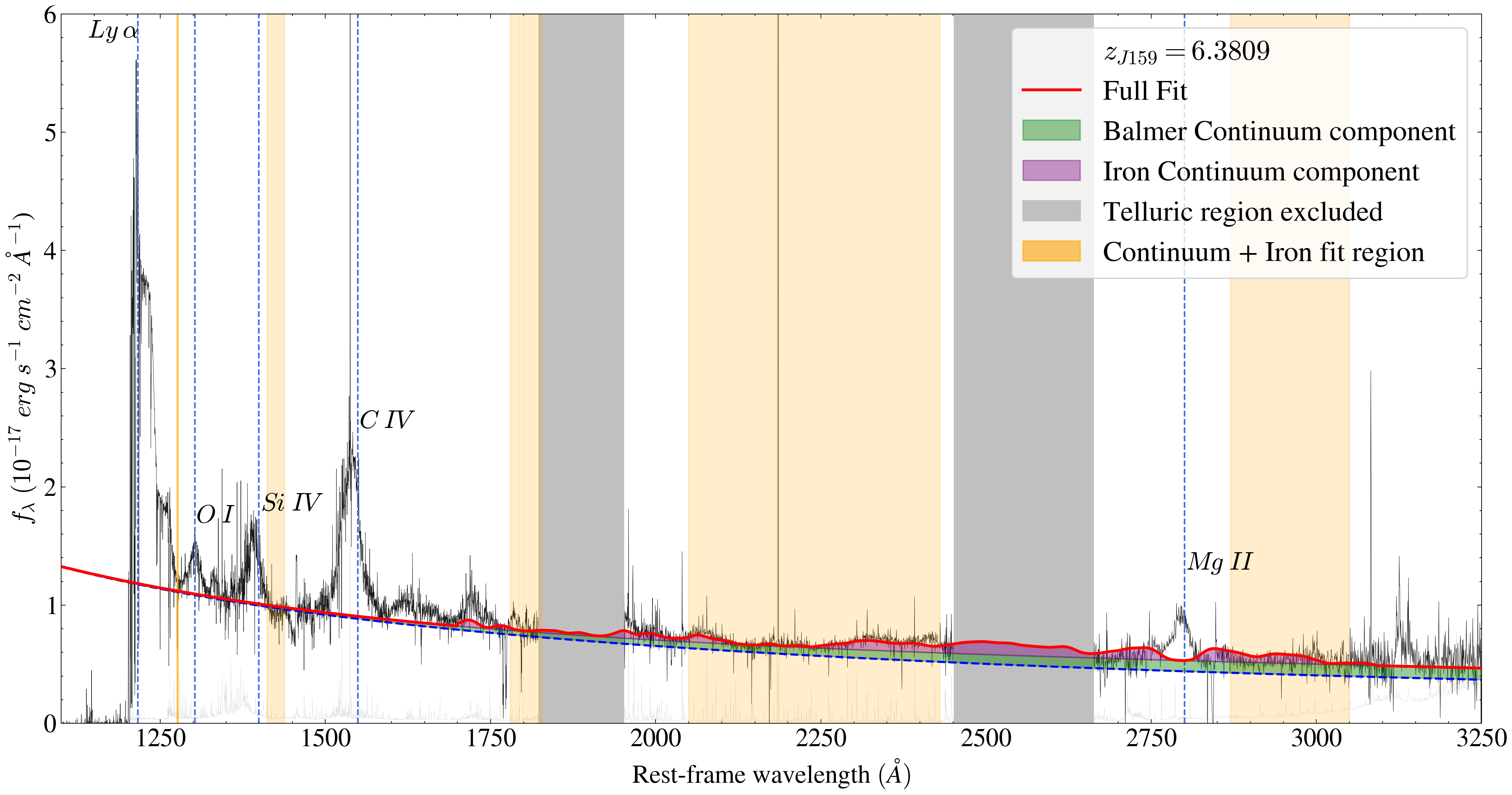} 
    \caption{Best fit to the PSO J159-02 X-Shooter spectrum. The spectrum, smoothed to 20 km s$^{-1}$, is depicted in black, with flux errors in gray. The dashed orange vertical bars indicate the wavelengths used for the fit of the continuum, being very careful to avoid spikes. The dashed gray vertical bars indicate the wavelength regions in which there are strong telluric residuals after the telluric correction. The red line denotes the combined fit of the continuum, where the blue dotted line indicates the power-law component, the green shaded area indicates the Balmer continuum contribution, and the shaded magenta area denotes the iron continuum contribution.}
    \label{J159_X-Shooter_spectrum}
\end{figure*}

Thus, we modeled the emission of the accretion disk as power-law normalized at 2500 \AA.
The Balmer pseudo-continuum component was modeled assuming partially optically thick gas clouds with uniform electron temperature, as discussed in \cite{2003_Dietrich}.
Here, we fix the normalization factor to 30\% of the power-law flux at the Balmer edge \citep{2020_Onoue}, and we consider an electron temperature of T$_e$ = 15000 K, typically used in the literature \citep{2003_Dietrich, Mazzucchelli_2017, 2020_Onoue}.
The iron pseudo-continuum beneath the \mgii\ broad emission line is modeled and fit with a template derived from the narrow-line Seyfert 1 galaxy I Zwicky-1 \citep{Vestergaard_2001}, linking the fitted iron Full Width Half Maximum (FWHM) to the fit of \mgii\ emission line.
Given the coupling between the iron emission and the \mgii\ line, we iteratively fit the continuum and the \mgii\ profile, updating the iron template parameters after each step. The procedure is repeated until the \mgii\ FWHM and redshift reach convergence \citep[see also][]{Shin_2019, 2020_S}.
The new full spectrum with all three components, smoothed to 20 km s$^{-1}$ using \texttt{Astrocook}\footnote{\url{https://das-oats.github.io/astrocook/}} \citep{2020_Astrocook, 2022_Astrocook}, is reported in Fig.~\ref{J159_X-Shooter_spectrum}.

From the continuum fit we derived some of the physical properties of our quasar, starting from power-law slope to flux densities and luminosities at 1350 \AA and 3000 \AA. We find a power-law slope of $\alpha_{\lambda} = - 1.19 \pm 0.02$.\\
Then, we follow \cite{Shen_2011} to estimate the bolometric luminosity:
\begin{equation}
    L_{bol} =  5.15 \times \lambda L_{\lambda,3000}
\end{equation}
obtaining $L_{bol} = (3.01 \pm 0.03) \times 10^{47}$ erg s$^{-1}$.

\subsubsection{Emission lines and black hole mass}
\begin{figure*}[ht!]
   \centering
   \begin{minipage}{0.49\hsize}
       \centering
       \includegraphics[width=\linewidth]{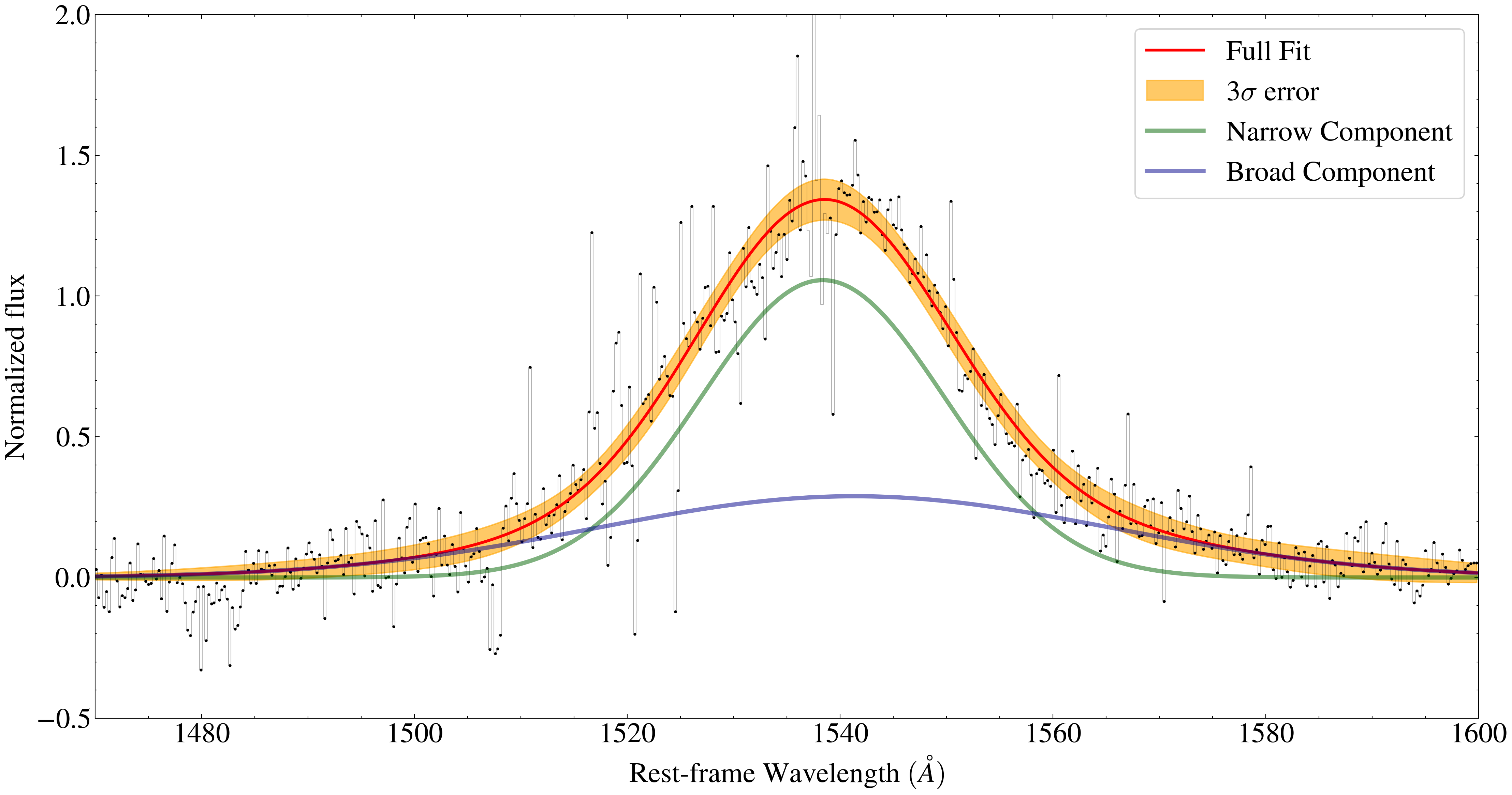}
   \end{minipage}
   \hfill
   \begin{minipage}{0.49\hsize}
       \centering
       \includegraphics[width=\linewidth]{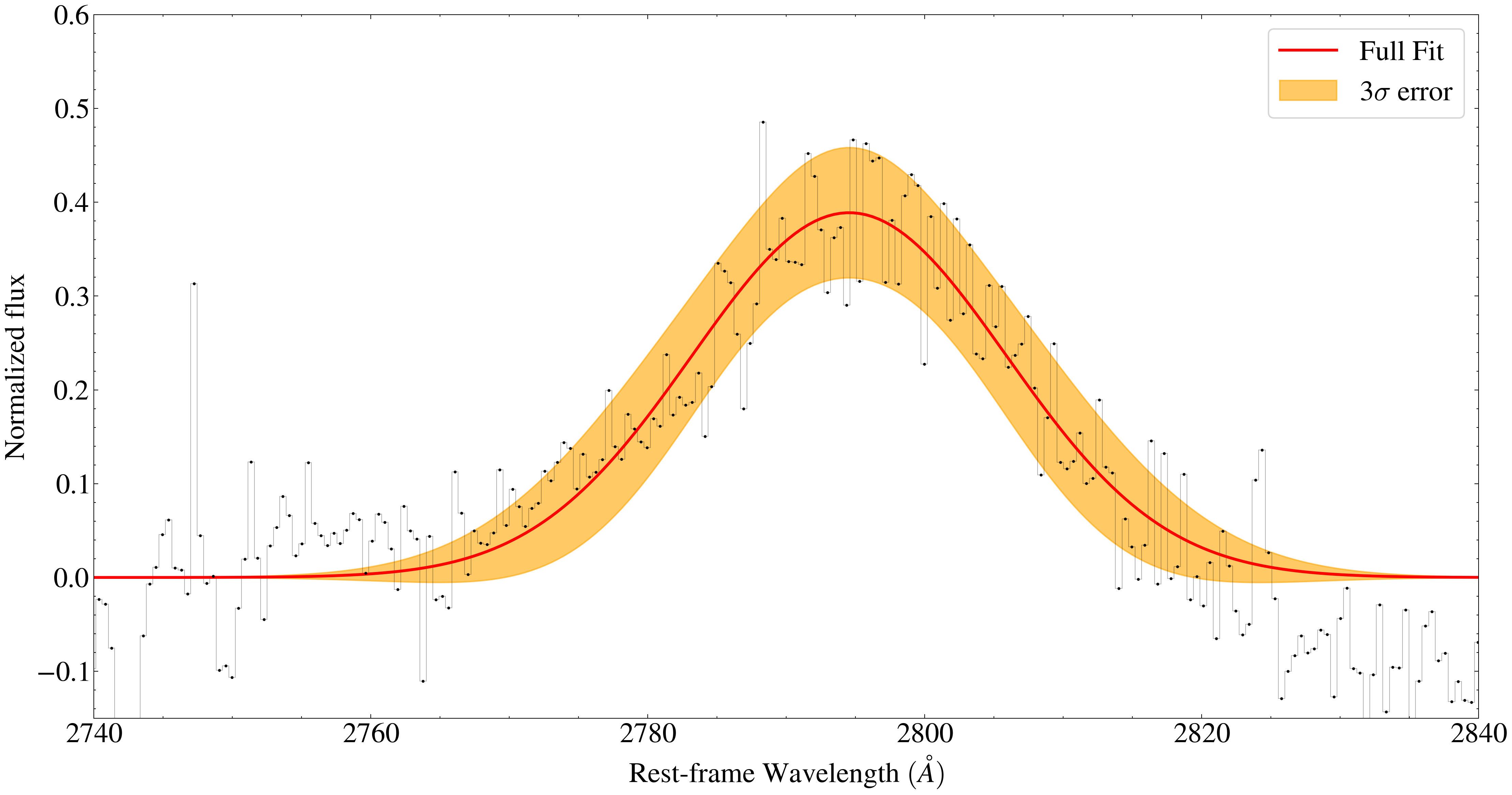}
   \end{minipage}
   \caption{Best-fit model of the \civ\ emission line (left-hand panel) and \mgii\ emission line (right-hand panel) in the rest-frame spectrum, rebinned to 50 km/s. The black points denotes the observed spectrum after subtracting continuum, while the red curve represents the total model given by double gaussian component. The green and blue curve denote the narrow and broad components, respectively. The orange shaded region marks the 3$\sigma$ uncertainty on the fit.}
   \label{CIV_MgII_em_lines}
\end{figure*}

\begin{table*}[ht!]
\tiny
\caption{Emission lines and quasar properties}                 
\label{table_emission_lines_quasar}    
\centering                        
\begin{tabular}{c c c c c c c}      
\hline\hline               
 & $z$ & FWHM$^a$ & EW & $\log(M_{\mathrm{BH}}/M_{\odot})$ & $\log(L_{\mathrm{Edd}}/\rm 10^{47}erg \ s^{-1} )$ & $\lambda_{\mathrm{Edd}}$\\
 & & \kms & \AA & & & \\
\hline                     
\civ\ & 6.3403 $\pm$ 0.0004 & 6104 $\pm$ 290 & 52.33 $\pm$ 1.22 & 9.53 $\pm$ 0.04 & 4.08 $\pm$ 0.04 & 0.74 $\pm$ 0.06 \\  
\mgii\ & 6.3685 $\pm$ 0.0002 & 2908 $\pm$ 146 & 21.17 $\pm$ 2.12 & 9.51 $\pm$ 0.05 & 4.24 $\pm$ 0.06 & 0.71 $\pm$ 0.05 \\  
\hline                               
\end{tabular}
\tablefoot{$^a$: We report the \civ\ FWHM not corrected from the outflow.}
\end{table*}
We modeled the \civ\ emission line as a double Gaussian, while the \mgii\ emission line, due to its low signal-to-noise, was modeled as a single Gaussian. The fits are reported in Fig.~\ref{CIV_MgII_em_lines}.

We calculate the peak wavelength from the maximum flux value of the full-line model (all components), the FWHM, the Equivalent Width (EW), the emission line luminosity, and finally we estimate the redshift of the quasar from the line. 
Once these quantities are obtained, we estimate the black hole mass. Assuming virial motion of the broad-line region (BLR) clouds around a supermassive black hole, the mass can be inferred from the virial theorem, which relates the gas velocity dispersion, traced by the FWHM of the broad emission line (FWHM$_{\rm BLR}$), to the gravitational potential of the black hole $M_{\rm BH}$:
\begin{equation}
    \mathrm{M_{BH}} = f \cdot \dfrac{\mathrm{R_{BLR}} \cdot \mathrm{FWHM^2_{BLR}}}{G}
\end{equation}
where R$_{\rm BLR}$ indicates the radius from the SMBH to the emission line, and $f$ is a factor that takes into account our poor knowledge about orientation, geometry and kinematics of the BLR.
Thus, to derive the black hole mass we used the so-called single-epoch virial relation provided by \cite{2006_Vestergaard} for the \civ\ emission line, and \cite{2009_Vestergaard} for the \mgii\ emission line. These are expressed as:
\begin{equation}
    \mathrm{M_{BH, \mgii}} = 10^{6.86} \Biggl[\dfrac{\mathrm{FWHM}_{\mgii}}{10^3 \; \mathrm{km \; s^{-1}}}  \Biggr]^2 \Biggl[ \dfrac{\lambda L_{\lambda}(3000 \mathring{\mathrm{A}})}{10^{44} \; \mathrm{erg \; s^{-1}}} \Biggr]^{0.5} M_{\odot}
\end{equation}
where $\lambda L_{\lambda}(3000 \mathring{\mathrm{A}}$) is the monochromatic luminosity at 3000 $\mathring{\mathrm{A}}$ rest-frame wavelength, and
\begin{equation}
     \mathrm{M_{BH, \civ}} = 10^{6.66} \Biggl[\dfrac{\mathrm{FWHM}_{\civ,corr}}{10^3 \; \mathrm{km \; s^{-1}}}  \Biggr]^2 \Biggl[ \dfrac{\lambda L_{\lambda}(1350 \mathring{\mathrm{A}})}{10^{44} \; \mathrm{erg \; s^{-1}}} \Biggr]^{0.53} M_{\odot}
\end{equation}
where $\lambda L_{\lambda}(1350 \mathring{\mathrm{A}}$) is the monochromatic luminosity at 1350 $\mathring{\mathrm{A}}$ rest-frame wavelength. Here, we adopt the outflow-corrected FWHM to account for the strong wind component affecting the \civ\ emission line \citep{Mazzucchelli_2023}. So, the expression for the corrected FWHM is taken from \cite{Coatman_2017}:
\begin{equation}
    \rm FWHM_{\civ\,corr} = \dfrac{FWHM_{\civ}}{0.36 \times \frac{\civ \ blueshift}{10^3 \ km \ s^{-1}} + 0.61 }
\end{equation}
where \civ\ blueshift is the velocity difference between the \civ\ centroid and the quasars’ systemic redshifts, obtained from [\cii] emission line \citep{2018_DeCarli}. Here the \civ\ blueshift corresponds to 1649 km/s.\\
From the black hole mass measurements, we estimate the Eddington luminosity as:
\begin{equation}
    L_{\mathrm{Edd, \civ/\mgii}} = 1.3 \times10^{38} \Biggl(\dfrac{M_{\rm BH, \civ/ \mgii }}{M_{\odot}} \Biggr) \; \mathrm{erg \; s^{-1}}
\end{equation}
and so, the Eddington ratio as:
\begin{equation}
    \lambda_{\mathrm{Edd, \civ/\mgii}} = \dfrac{L_{\mathrm{bol}}}{L_{\mathrm{Edd, \civ/\mgii}}} .
\end{equation}
All the quantities obtained from the \civ\ and \mgii\ emission lines are listed in Table \ref{table_emission_lines_quasar}.

\section{PSO J159-02 absorption systems}
\label{Appendix_C}
In this appendix, we present the five absorption systems found in PSO J159-02. The identified ionic transitions are \oi\ $\lambda\, 1302$ \AA, \cii\ $\lambda\lambda\, 1334, 1335$ \AA, \siii\ $\lambda\lambda\lambda\, 1260, 1304, 1526$ \AA, \feii\ $\lambda\lambda\lambda\lambda\lambda\lambda\,, 1608, 2344, 2374, 2382, 2586, 2600$ \AA, \alii\ $\lambda\, 1670$ \AA, \mgii\ $\lambda\lambda\, 2796, 2803$ \AA. 
All absorption lines were fitted with Voigt profiles using the analysis software \texttt{Astrocook} \citep{2020_Astrocook}. The results of the fit are reported in Table \ref{Table_Voigt_Fits}.  
\begin{table}[ht!]
\tiny
\caption{Voigt profile fit results for the absorption systems identified towards PSO J159-02.} 
\label{Table_Voigt_Fits}
\centering
\begin{tabular}{l c c c}
\hline\hline
Ion & $z_{\rm comp}$ & $b$ & $\log N$ \\
 & & \kms & (cm$^{-2}$) \\
\hline
\multicolumn{4}{c}{$z_{\rm abs} = 5.73484$} \\
\hline
\cii\ 1334  & $5.73316 \pm 0.00001$ & $6.00$             & $13.266 \pm 0.023$ \\
            & $5.73491 \pm 0.00001$ & $24.45 \pm 0.32$   & $14.642 \pm 0.012$ \\
\cii* 1335$^a$  & $5.73491$              & $6.00$             & $13.669 \pm 0.021$ \\
\siii\ 1526$^a$ & $5.73491$              & $27.73 \pm 2.49$   & $14.195 \pm 0.034$ \\
\feii\ 2600$^a$ & $5.73491$              & $9.00$             & $14.261 \pm 0.020$ \\
\feii\ 1608$^{a,b}$ & $5.73491$              & $9.00$             & $14.261$ \\
\feii\ 2344$^{a,b}$ & $5.73491$              & $9.00$             & $14.261$ \\
\feii\ 2374$^{a,b}$ & $5.73491$              & $9.00$             & $14.261$ \\
\feii\ 2382$^{a,b}$ & $5.73491$              & $9.00$             & $14.261$ \\
\feii\ 2586$^{a,b}$ & $5.73491$              & $9.00$             & $14.261$ \\
\alii\ 1670$^a$ & $5.73491$              & $9.00$             & $13.400 \pm 0.075$ \\
\mgii\ 2796$^a$ & $5.73491$              & $9.00$             & $15.461 \pm 0.137$ \\
\mgii\ 2803$^a$ & $5.73491$              & $9.00$             & $15.461$ \\
\hline

\multicolumn{4}{c}{$z_{\rm abs} = 5.91269$} \\
\hline
\oi\ 1302  & $5.91166 \pm 0.00002$ & $6.00$             & $13.552 \pm 0.036$ \\
           & $5.91308 \pm 0.00001$ & $39.08 \pm 0.47$   & $14.410 \pm 0.008$ \\
           & $5.9154 \pm 0.0001$ & $6.00$             & $13.284 \pm 0.093$ \\
\siii\ 1304$^d$ & $5.91166$              & $6.00$             & $13.142 \pm 0.046$ \\
           & $5.91308$              & $32.87 \pm 0.17$   & $13.656 \pm 0.016$ \\
\siii\ 1526$^d$ & $5.91308$              & $32.87$            & $13.656$ \\
\cii\ 1334$^d$ & $5.91166$              & $16.13 \pm 0.30$   & $13.904 \pm 0.021$ \\
           & $5.91308$              & $29.78 \pm 0.45$   & $14.265 \pm 0.012$ \\
\feii\ 2374$^d$ & $5.91308$              & $15.00$            & $13.023 \pm 0.024$ \\
\feii\ 1608$^{c,d}$ & $5.91308$              & $15.00$            & $13.023$ \\
\feii\ 2344$^{c,d}$ & $5.91308$              & $15.00$            & $13.023$ \\
\feii\ 2382$^{c,d}$ & $5.91308$              & $15.00$            & $13.023$ \\
\feii\ 2600$^{c,d}$ & $5.91308$              & $15.00$            & $13.023$ \\
\hline

\multicolumn{4}{c}{$z_{\rm abs} = 5.92243$} \\
\hline
\cii\ 1334  & $5.92315$             & $38.52 \pm 5.47$   & $13.347 \pm 0.050$ \\
\siii\ 1304$^a$ & $5.92168 \pm 0.00007$ & $13.06 \pm 5.27$   & $12.982 \pm 0.098$ \\
            & $5.92315 \pm 0.00006$ & $9.00$             & $12.976 \pm 0.062$ \\
\hline

\multicolumn{4}{c}{$z_{\rm abs} = 6.0554$} \\
\hline
\oi\ 1302   & $6.05445 \pm 0.00001$ & $18.65 \pm 0.79$   & $14.434 \pm 0.011$ \\
           & $6.05623 \pm 0.00003$ & $6.00$             & $13.300 \pm 0.087$ \\
\siii\ 1304$^d$ & $6.05445$              & $21.39 \pm 5.76$   & $13.223 \pm 0.063$ \\
\cii\ 1334$^d$  & $6.05445$              & $32.97 \pm 3.64$   & $13.948 \pm 0.037$ \\
           & $6.05623$              & $40.36 \pm 3.38$   & $14.016 \pm 0.029$ \\
\mgii\ 2796$^d$ & $6.05445$              & $9.00$             & $12.574 \pm 0.131$ \\
           & $6.05520 \pm 0.00004$ & $9.00$             & $13.187 \pm 0.078$ \\
           & $6.05623$              & $9.00$             & $13.145 \pm 0.087$ \\
\mgii\ 2803$^d$ & $6.05445$              & $9.00$             & $12.574$ \\
           & $6.05520$              & $9.00$             & $13.187$ \\
           & $6.05623$              & $9.00$             & $13.145$ \\
\hline

\multicolumn{4}{c}{$z_{\rm abs} = 6.21904$} \\
\hline
\cii\ 1334  & $6.21829 \pm 0.00018$ & $19.94 \pm 13.46$  & $12.998 \pm 0.152$ \\
           & $6.21977$              & $26.11 \pm 9.62$   & $13.219 \pm 0.098$ \\
\siii\ 1260$^a$ & $6.21977 \pm 0.00002$ & $14.11 \pm 1.83$   & $12.357 \pm 0.019$ \\
\hline

\multicolumn{4}{c}{$z_{\rm abs} = 6.2385$} \\
\hline
\cii\ 1334  & $6.23695$              & $21.15 \pm 3.35$   & $13.836 \pm 0.031$ \\
           & $6.24017 \pm 0.00009$ & $47.16 \pm 4.86$   & $13.829 \pm 0.036$ \\
\siii\ 1260$^a$ & $6.23695 \pm 0.00001$ & $21.42 \pm 0.84$   & $12.865 \pm 0.010$ \\
           & $6.23951 \pm 0.00002$ & $45.02 \pm 1.57$   & $12.822 \pm 0.011$ \\
\mgii\ 2796$^a$ & $6.23695$              & $9.00$             & $12.895 \pm 0.059$ \\
           & $6.23951$              & $9.00$             & $13.048 \pm 0.052$ \\
\mgii\ 2803$^a$ & $6.23695$              & $9.00$             & $12.895$ \\
           & $6.23951$              & $9.00$             & $13.048$ \\
\hline
\end{tabular}
\tablefoot{List of the individual kinematic components fitted for each ionic transition. Values without explicit errors are tied to other components during the fitting process. For those components for which the b values go below the specific X-Shooter arm resolution, we set their values to the minimum possible: $6.00$ and $9.00$ for the VIS and NIR arms, respectively.\\
$^a$ The z-values for this component were tied to those for \cii.\\
$^b$ The b- and $\log(N)$-values for this component were tied to those for \feii\ 2600.\\
$^c$ The b- and $\log(N)$-values for this component were tied to those for \feii\ 2374.\\
$^d$ The z-values for this component were tied to those for \oi.}
\end{table}

\begin{figure}[h!]
    \includegraphics[width=\linewidth]{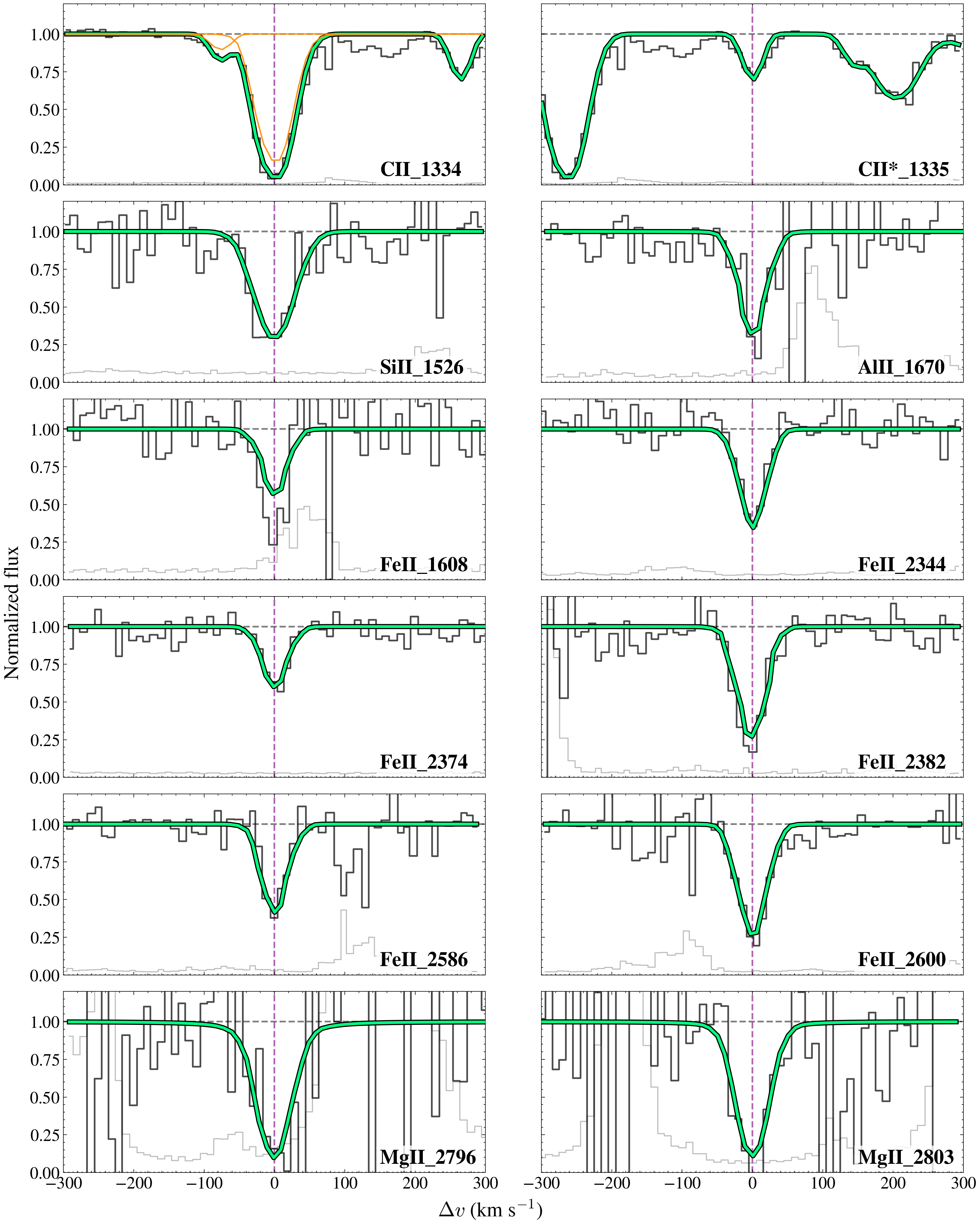}
    \caption{Voigt profile fits of absorption system at $z = 5.73484$ found in J159 quasar spectrum. The flux is denoted in black, the error in gray, and full Voigt profile in green. The individual Voigt component fits are shown in orange.}
    \label{CII_Abs_j159_1}
\end{figure}
\begin{figure}[h!]
    \includegraphics[width=\linewidth]{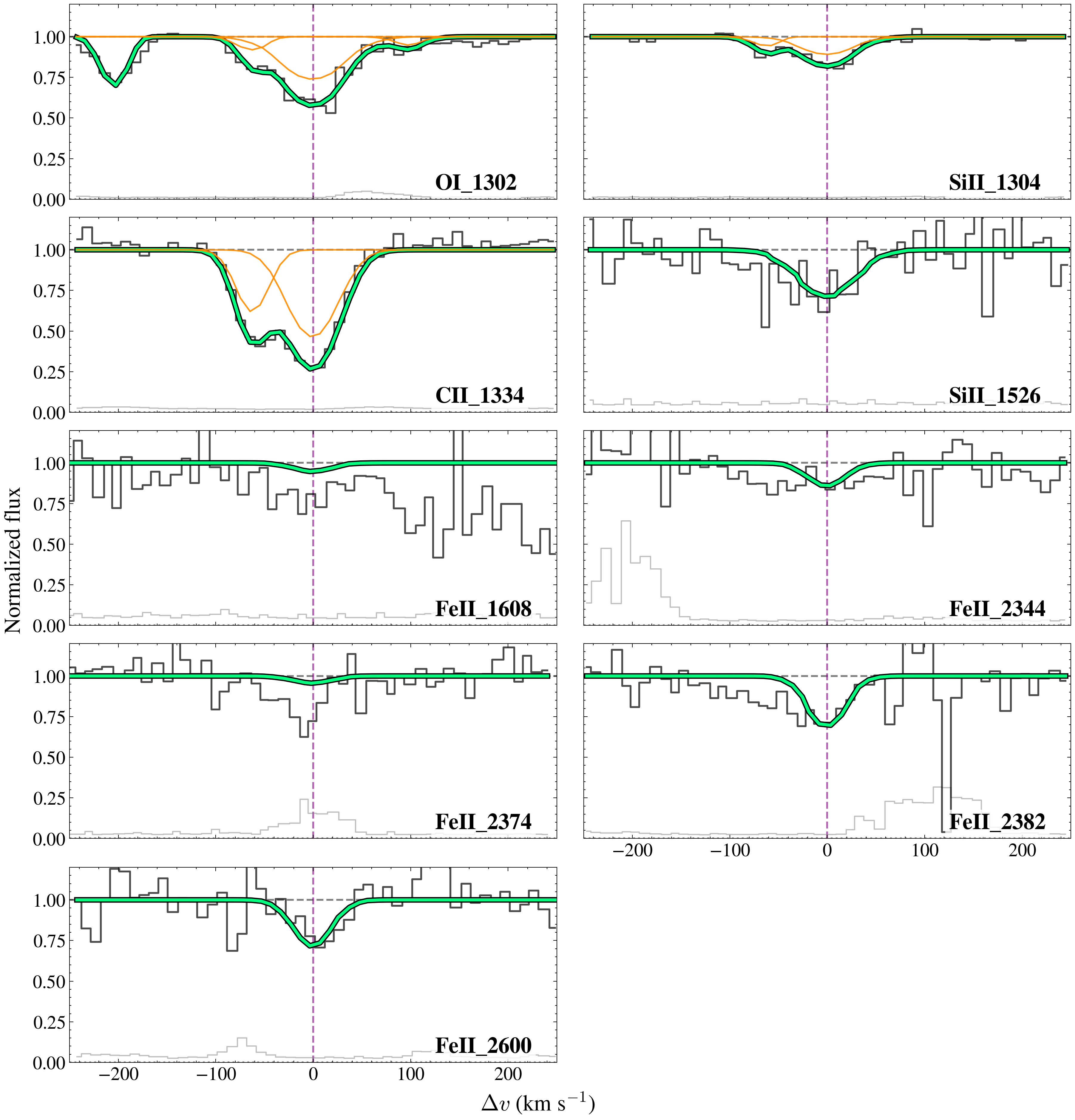} 
    \caption{Voigt profile fits of absorption system at $z = 5.91269$ found in J159 quasar spectrum.}
\end{figure}
\begin{figure}[h!]
    \includegraphics[width=\linewidth]{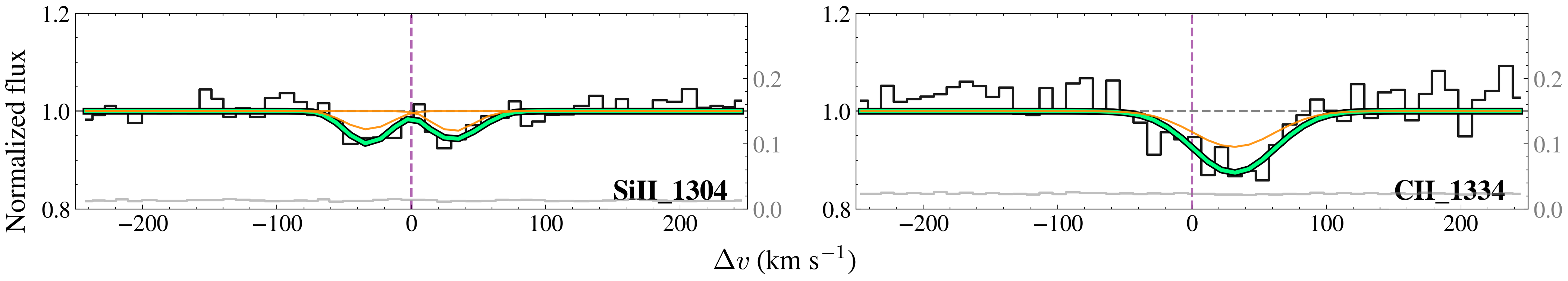}
    \caption{Voigt profile fits of absorption system at $z = 5.92243$ found in J159 quasar spectrum.}
\end{figure}
\begin{figure}[h!]
    \includegraphics[width=\linewidth]{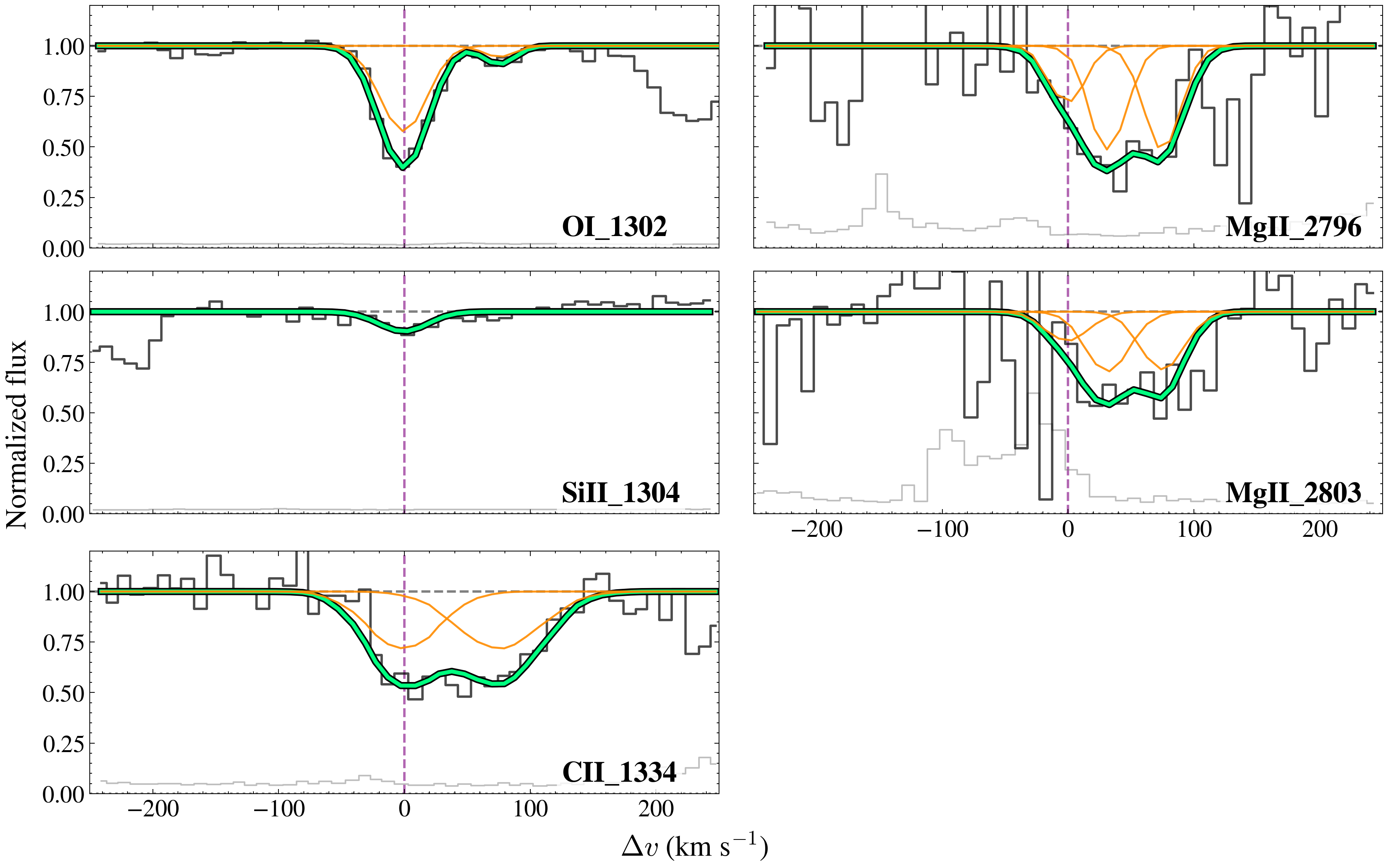} 
    \caption{Voigt profile fits of absorption system at $z = 6.0554$ found in J159 quasar spectrum.}
\end{figure}
\begin{figure}[h!]
    \includegraphics[width=\linewidth]{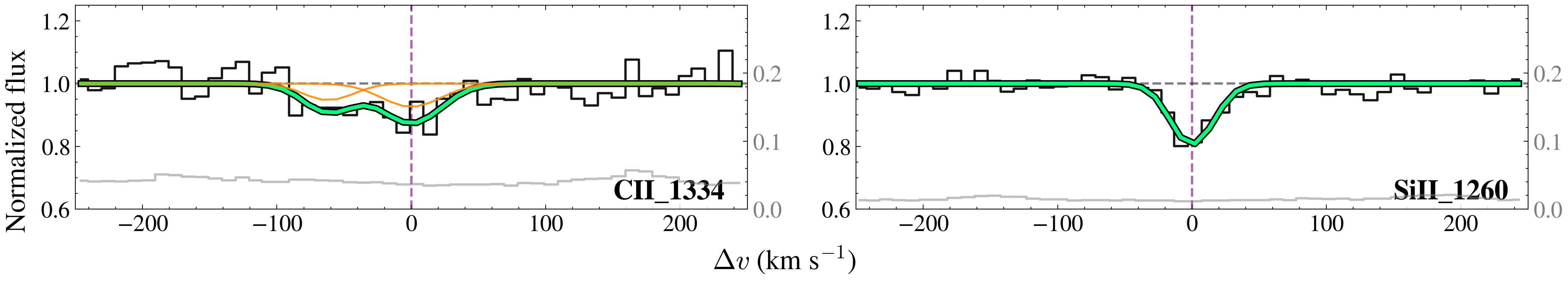} 
    \caption{Voigt profile fits of absorption system at $z = 6.21904$ found in J159 quasar spectrum.}
\end{figure}
\begin{figure}[h!]
    \includegraphics[width=\linewidth]{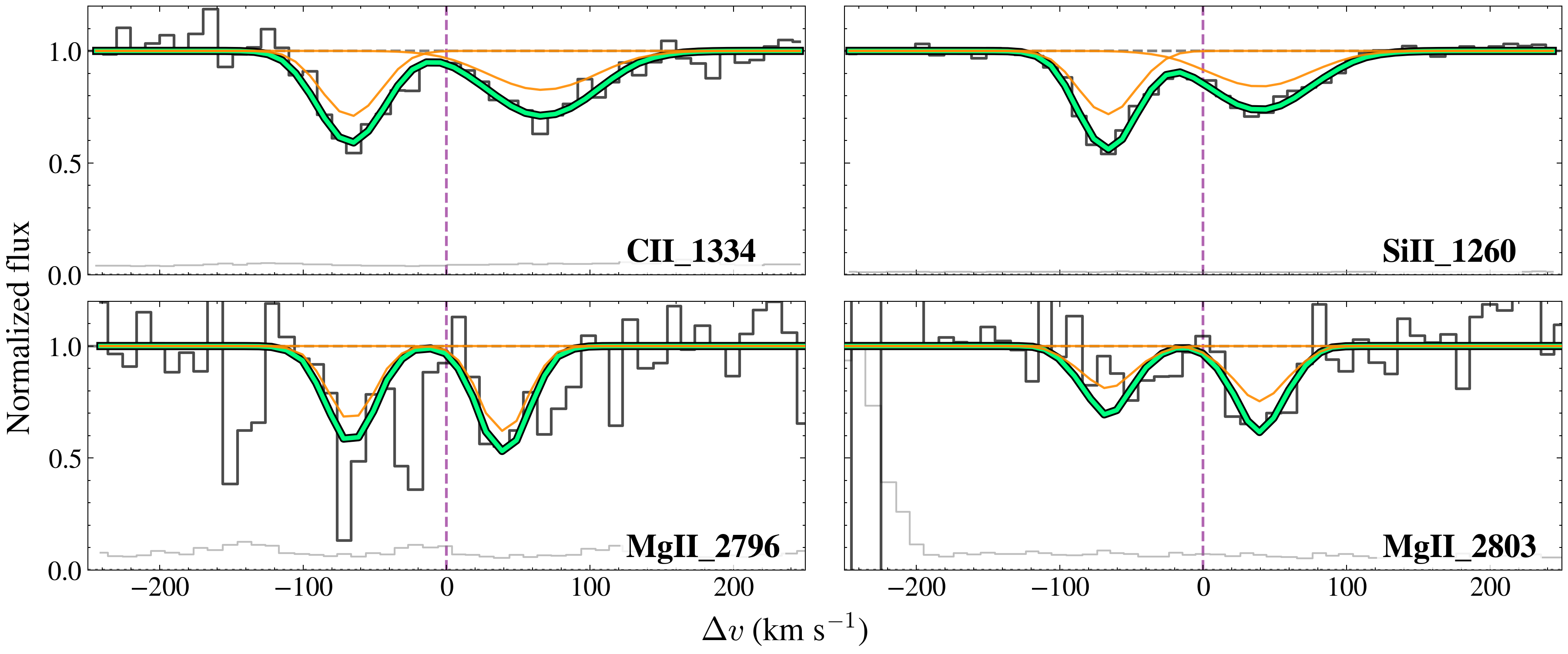}
    \caption{Voigt profile fits of absorption system at $z = 6.2385$ found in J159 quasar spectrum.}
    \label{CII_Abs_j159_1}
\end{figure}
\end{appendix}
\end{document}